\documentclass[12pt]{article}
\pdfoutput=1
\usepackage{putex}
\usepackage{feyn}
\usepackage{xcolor}

\usepackage{graphicx}
\usepackage{epstopdf}
\usepackage{enumerate}
\usepackage{cite}
\usepackage{tensor}
\usepackage{slashed}
\usepackage{feynmf}

\usepackage{hyperref}

\numberwithin{equation}{section}

\newcommand{\abs}[1]{\left\lvert #1 \right\rvert}

\newcommand {\be} {\begin {equation}}
\newcommand {\ee} {\end {equation}}

\newcommand {\bes} {\begin {equation*}}
\newcommand {\ees} {\end {equation*}}

\newcommand{\es}[2] {\begin{equation} \label{#1} \begin{split} #2 \end{split} \end{equation}}

\def\CL{{\cal L}}

\newcommand{\beq}{\begin{equation}}
\newcommand{\eeq}{\end{equation}}

\newcommand{\spc}[2][c]{\begin{tabular}[#1]{@{}c@{}}#2\end{tabular}}

\begin{document}

\preprint{PUPT-2434}

\institution{PU}{Department of Physics, Princeton University, Princeton, NJ 08544}
\institution{PCTS}{Princeton Center for Theoretical Science, Princeton University, Princeton, NJ 08544}

\title{
A Crack in
the Conformal Window
}

\authors{Benjamin R.~Safdi,\worksat{\PU} Igor R.~Klebanov\worksat{\PU,\PCTS} and Jeongseog Lee\worksat{\PU}
}

\abstract{
In ${\cal N}=2$ superconformal three-dimensional field theory the R-symmetry is determined by
locally maximizing the free energy $F$ on the three-sphere. Using $F$-maximization, we
study the ${\cal N}=2$ supersymmetric $U(N_c)$ gauge theory coupled to $N_f$ pairs of
fundamental and anti-fundamental superfields in the Veneziano large $N_c$ limit, where $x=N_f/N_c$ is kept fixed.
This theory has a superconformal window $1 \leq x \leq \infty$, while for $x<1$ supersymmetry is broken.
As we reduce $x$ we find ``a crack in the superconformal window'' -- a critical value
$x_c\approx 1.45$ where the monopole operators reach the unitarity bound. To continue the theory to $x<x_c$ we assume that the
monopoles become free fields, leading to an accidental global symmetry. Using the Aharony dual description of the theory for $x<x_c$
allows us to determine the R-charges and $F$ for $1\leq x < x_c$. Adding a Chern-Simons term removes the transition at $x_c$.
In these more general theories we study the scaling dimensions of meson operators as functions of $x$ and $\kappa=\abs{k}/N_c$. We find that there is an interesting transition in behavior at $\kappa=1$.
 }

\date{January 2013}

\maketitle

\tableofcontents

\section{Introduction and Summary}

Supersymmetry is a powerful tool for understanding strongly coupled dynamics in quantum field theory. In four dimensions the infrared behavior of flavored ${\cal N} = 1$ supersymmetric QCD (SQCD) was understood~\cite{Seiberg:1994bz,Seiberg:1994pq,Intriligator:1995id,Intriligator:1995ne} in the 1990's, and many similarities were found with the expected behavior of non-supersymmetric QCD.  The supersymmetric theories have the advantage of being much more tractable at strong coupling than their non-supersymmetric cousins. For example, the SQCD theory with $SU(N_c)$ gauge group and $N_f$ non-chiral massless flavors, where
each flavor multiplet consists of a fundamental and anti-fundamental chiral superfield, $(Q, \tilde Q)$,
 flows to an interacting infrared fixed point in the Seiberg conformal window ${3 N_c \over 2} < N_f < 3 N_c$\cite{Seiberg:1994pq}. A similar conformal window is believed to exist for the non-supersymmetric $SU(N_c)$
gauge theory coupled to massless flavors \cite{Banks:1981nn}. Its upper boundary, $N_f=11 N_c/2$, is determined by asymptotic freedom, but its lower boundary is not yet known precisely.

The dynamics of gauge theories in three dimensions is of much interest due, in part, to their relation to
statistical mechanics and condensed matter physics.
It is expected that the $U(N_c)$ gauge theory with $N_f$ massless flavors flows to an interacting infrared fixed point when $N_f > N_{\text{crit}}$, where $N_{\text{crit}}$ is some critical number of flavors~\cite{PhysRevD.33.3704,1988PhRvL..60.2575A}.
 For large $N_f$ the scaling dimensions of composite operators may be calculated using the $1/N_f$ expansion. For $N_f < N_{\text{crit}}$ the theory is thought to flow to a gapped phase in the infrared,
 though
 this phenomenon is difficult to study
 due to the strong coupling nature of the transition.

In order to gain more insight into the nature of the conformal window in 3-d gauge theories,
it is instructive to study such theories with ${\cal N} = 2$ supersymmetry, which are under an improved theoretical control. The $U(N_c)$ gauge theory with $N_f$ non-chiral flavors
flows to an IR fixed point for
$N_f\geq N_c$, while for $N_f < N_c$ supersymmetry is broken \cite{Aharony:1997bx}.
 For $N_f\geq N_c$ the ${\cal N}=2$ superconformal theories possess the Aharony duality~\cite{Aharony:1997gp} mapping them to $U(N_f-N_c)$ theories with $N_f$ non-chiral flavors along with extra neutral matter, in analogy with the Seiberg duality~\cite{Seiberg:1994pq} in 4-d ${\cal N}=1$ theories. When a Chern-Simons term is added, this is generalized to the
Giveon-Kutasov duality~\cite{Giveon:2008zn}.

During the past three years many new insights into the 3-d ${\cal N}=2$ theories have been obtained using the method
of localization on the three-sphere \cite{Kapustin:2009kz,Jafferis:2010un,Hama:2010av}.
The matter fields in ${\cal N} = 2$ theories have non-trivial anomalous dimensions at conformal fixed points, whereas the anomalous dimensions at fixed points in theories with more supersymmetry vanish.
The R-symmetry in ${\cal N} = 2$ theories is abelian, and
it may mix with other abelian symmetries in the infrared.
The correct R-symmetry at the IR fixed point may be calculated using the principle of $F$-maximization~\cite{Jafferis:2010un,Jafferis:2011zi,Closset:2012vg}, which states that the correct R-symmetry locally maximizes the Euclidean three-sphere free energy.

The calculations on $S^3$ have also led to precise checks \cite{Willett:2011gp,Bashkirov:2011vy,Benini:2011mf} of the Aharony and Giveon-Kutasov dualities.  These dualities, which are similar to the Seiberg duality~\cite{Seiberg:1994pq} in four dimensions, provide dual (magnetic) theories which are thought to flow to the same infrared fixed points as the original (electric) theories.  The authors of~\cite{Willett:2011gp} showed that the $S^3$ partition functions at the IR fixed points of the electric and magnetic theories agree when treated as analytic functions of the trial R-charges, providing evidence that the fixed points are indeed equivalent.
However, one feature of the Aharony duality does not seem to have been fully clarified, and this is one of the subjects of this paper. In the theories without Chern-Simons level it was observed~\cite{Willett:2011gp} that the partition functions of the $U(N_c)$ electric theories fail to converge for a small enough number of flavors when the R-charges are treated as real variables.
In the Veneziano limit~\cite{Veneziano:1976wm}, where $N_c$ is taken to infinity while keeping the ratio $x = {N_f \over N_c}$ fixed, we find that this divergence occurs at the critical value $x = x_c$, which we determine numerically to be $x_c \approx 1.45$.  For $x < x_c$ a new ``accidental'' global symmetry emerges in the infrared
and mixes with the IR R-symmetry.  A key insight into the nature of this global symmetry is found by studying the scaling dimension of the protected monopole operators, which are local operators in three-dimensions, as functions of $x$.  We find that the monopole operators are above the unitarity bound for $x > x_c$ and reach the unitarity bound $\Delta=1/2$ at $x = x_c$.  When $x < x_c$ the monopole fields become
free and decouple, so that the accidental global symmetry acts on the monopole sector.\footnote{We are grateful to O. Aharony and I. Yaakov
  for suggesting this possibility to us.}  While this global symmetry acts on the monopole fields in the electric theory, in the Aharony dual theory it acts on two chiral superfields that don't emerge from the gauge sector.  This allows us to set the R-charges of these two superfields to $1/2$ and then use $F$-maximization in the magnetic theory to calculate the $S^3$ free energy and the R-symmetry of the IR fixed point in the theories with $x < x_c$. This procedure is in line with the general approach to handling the accidental symmetries
  proposed in \cite{Morita:2011cs,Agarwal:2012wd}, which is inspired by the work \cite{Kutasov:2003iy,Barnes:2004jj} on accidental symmetries in 4-d ${\cal N}=1$ gauge theories.
We refer
to the transition in the behavior of the theory at $x=x_c$ as ``a crack in the superconformal window.'' The
existence of the ``crack'' has interesting effects on the properties of observables. For example, as we show in section 2.5,
when $F$ is plotted at fixed $N_f$ as a function of $N_c$, it is peaked at the ``crack.''

We recall that, in the 4-d ${\cal N}=1$ theories,
 as $x$ is decreased the dimensions of the
meson operators $\tilde Q^a Q_b$ decrease and eventually reach the unitarity bound at the lower edge of the conformal window, $x=3/2$
\cite{Seiberg:1994bz}.
A similar phenomenon occurs in 3-d, with meson operators reaching the unitarity bound at $x=1$. However, this is not the first transition
that affects the 3-d theories as $x$ is decreased. The monopole operators, which are special to the 3-d case, reach the unitarity bound {\it before} the meson operators do. The fact that the monopoles are free for $1\leq x<1.45$ makes the 3-d ${\cal N}=2$ theory in this range reminiscent
of the free magnetic phase found for $1\leq x <3/2$ in
the 4-d ${\cal N}=1$ $SU(N_c)$ gauge theory \cite{Seiberg:1994pq}. However, the 3-d theory is not free for $1\leq x<1.45$: in addition to the free monopoles it includes an interacting superconformal sector.

The key role of the monopole operators in bringing about the transition at $x = x_c$ is also reminiscent of the Polyakov mechanism~\cite{Polyakov:1976fu} for confinement in three dimensions. This leads us to raise the question of whether in the non-supersymmetric versions of these theories the monopole operators also reach the unitarity bound before the meson operators do.
 Since non-supersymmetric theories, as far as we know, do not posses anything similar to the
 Aharony duality, we conjecture that the theory is not conformal for $x$ below the value where
 the monopoles saturate the unitarity bound (i.e. the ``crack'' we have found in the superconformal window is analogous to the edge
 of the non-supersymmetric conformal window).
 It is thus tempting to conjecture that, at the lower edge of the 3-d non-supersymmetric conformal window for $U(N_c)$
 theories coupled to massless flavors, the monopole operators reach the unitarity
 bound while the meson operators, such as $\bar \psi_a \psi_b$ in 3-d QCD, are still above the unitarity bound.\footnote{For discussions of other possibilities for the physics at the edge of conformal window, see~\cite{PhysRevD.33.3704,1988PhRvL..60.2575A,Kaplan:2009kr}.}

In the ${\cal N}=2$ theories, the transition at $x = x_c$ disappears when we add in a Chern-Simons (CS) term at level $k$ to the gauge sector.  This is because there are no gauge invariant BPS operators which can be constructed from the monopole operators in the theories with $k \neq 0$.  At the level of the $S^3$ partition function, the partition function is seen to converge everywhere above the supersymmetry bound when the CS level is non-vanishing.  It is still instructive in this case to keep track of the scaling dimensions of the protected meson operators, which are constructed from the flavors, as functions of $x = {N_f \over N_c}$ and $\kappa = {\abs{k} \over N_c}$ in the Veneziano limit.  We find three different types of behavior at small $x$ depending on whether $\kappa < 1$, $\kappa = 1$, or $\kappa > 1$.  The theories with $\kappa < 1$ reach the supersymmetry bound at $x = 1 - \kappa$, and at this point the meson operators have dimension $1/2$ and become free fields.  The theories with $\kappa = 1$ are quite special; at small $x$
the scaling dimensions of the meson operators approach ${2 \over 3}$ due to the cubic superpotential in the magnetic Giveon-Kutasov theory.  In the theories with $\kappa > 1$ the meson dimensions approach unity at small $x$, which is likely due to an enhanced higher spin symmetry in this limit~\cite{Aharony:2011jz,Giombi:2011kc,Chang:2012kt}.

\section{Flavored ${\cal N}=2$ gauge theory without Chern-Simons term}

In this section we consider the non-chiral ${\cal N} = 2$ theory with gauge group $U(N_c)$ at vanishing Chern-Simons level and with $N_f$ non-chiral flavor multiplets $(Q_a, \tilde Q^a)$, $a=1, \ldots, N_f$. This theory has a supersymmetric vacuum if $N_f \geq N_c$ \cite{Aharony:1997bx}.

\subsection{Global symmetries and monopole operators}

Naively the theory has a $U(N_f) \times U(N_f)$ global flavor symmetry.  However, the diagonal $U(1)_V$, which rotates the $Q_a$ and $\tilde Q^a$ by opposite phases, is gauged, and this reduces the global flavor symmetry to $SU(N_f) \times SU(N_f) \times U(1)_A$, where the $U(1)_A$ rotates the two chiral superfields by the same phase.  There is also a $U(1)_R$ symmetry.  At superconformal fixed points the scaling dimension of an operator is equal to the absolute value of its $U(1)_R$ charge.  The UV R-charges of the chiral superfields $(Q_a, \tilde Q^a)$ thus take the free value $1/2$.  The correct R-symmetry at the IR fixed point may be a combination of the UV R-symmetry and the other global $U(1)$ symmetries.

There are also $N_c$  topological currents $j_{\text{top}} = \star \tr F$, where the field strength $F$ is proportional to a Cartan generator of $U(N_c)$.  The monopole operators, which are local operators in three-dimensions, are charged under the topological $U(1)$'s.
 More specifically, in the presence of a monopole operator charged under the topological $U(1)$ with current $j_{\text{top}} = \star \tr F$ inserted at the origin, the field strength takes the form
\es{F1}{
F = {M \over 2} \star d {1 \over \abs{x}} \,,
}
where $M$ is an element of the Cartan subalgebra.  The Dirac quantization condition restricts
\es{Mparam}{
M = \text{diag} ( q_1, \dots, q_{N_c} ) \,,
}
 with integer $q_1 \geq q_2 \dots \geq q_{{N_c}}$,
 up to gauge transformations.  Semi-classically we may construct BPS field configurations in ${\cal N} = 2$ theories by combining the field strength in~\eqref{F1} with background configurations for the adjoint scalar $\sigma$ in the vector multiplet.  The monopole can then be thought of as being the spin-$0$ component of a chiral superfield.  These chiral superfields parameterize the classical Coulomb branch of the theory.
However, in the quantum theory only two of these monopole operators survive in the chiral ring,  and these are the monopole operators with Cartan generators $M = \text{diag} ( \pm 1, 0, \dots, 0 ) $~\cite{Aharony:1997bx,deBoer:1997kr}.  We refer to these special monopole operators which remain in the chiral ring as $V_+$ and $V_-$.  A summary of the global and gauge symmetries of the theory is given in table~\ref{EMTable}.
\begin{table}
\begin{center}
\begin{tabular}{| c | ccccc |}
 \hline
  Chiral Field  & $U(N_c)$ & $SU(N_f) \times SU(N_f)$ & $U(1)_A$ & $U(1)_J$ &  $U(1)_{R-\text{UV}}$ \\
   \cline{1-6} $Q_a$ &  $N_c$ &  $(N_f , 1)$  & $1$ & $0$ & ${1 \over 2}$ \\
  $\tilde Q^a $ &  $\overline{N_c}$ &  $(1 , \overline{N_f})$  & $1$ & $0$ & ${1 \over 2}$ \\
 $ V_{\pm}$ & $1$ & $(1,1)$ & $-N_f$ & $\pm 1$ & $ {N_f \over 2} - N_c + 1$ \\
  \hline
\end{tabular}
\end{center}
\caption{ The chiral superfields which generate the chiral ring of $U(N_c)$ ${\cal N} = 2$ SYM theory without Chern-Simons level along with their charges under the gauge and global symmetries.  Note that the $U(1)_R$ symmetry stated is that of the UV fixed point.  The correct R-symmetry in the IR will be a mixture of $U(1)_{R - \text{UV}}$ and $U(1)_A$.}
\label{EMTable}
\end{table}

 For our purposes the most important property of the monopoles is their IR R-charge.
 The monopoles acquire an R-charge
 \cite{Borokhov:2002cg,Gaiotto:2008ak,Jafferis:2009th}, with the result
\es{monopoleRcharge}{
\Delta_{V_{\pm}} =  - N_c + 1 + N_f (1-\Delta)  \,,
}
where the first contribution is from the gauginos and the second is from the fermions in the flavor multiplets.  Here $\Delta$ is the IR R-charge of the flavor supermultiplets.  In the UV $\Delta = 1/2$, and this gives the value for $\Delta_{V_{\pm}}$ in table~\ref{EMTable}.  The monopole operators also acquire a $U(1)_A$ charge of $-N_f$ at one loop.

One might worry that~\eqref{monopoleRcharge} is not exact since the topological $U(1)_J$ under which the monopole operators are charged can in principle mix with the R-symmetry.  To understand the resolution to this question, it is useful to review the relevant discrete symmetries of the theory.  The SYM theory is invariant under charge conjugation symmetry and parity symmetry.  Charge conjugation acts by exchanging the fundamental flavors with the anti-fundamental flavors.  Parity symmetry acts by exchanging the monopole operators $V_+$ and $V_-$.  Thus, charge conjugation symmetry implies that the R-charge of the fundamental flavors equals the R-charge of the anti-fundamental flavors, and parity symmetry restrict the R-charges of $V_+$ and $V_-$ to be the same.  The R-symmetry cannot mix with $U(1)_J$ since this would necessarily lead to different R-charges for the two monopole operators.

\subsection{Monopole scaling dimensions and the unitarity bound}

 We must have $\Delta_{V_\pm} \geq \frac12$ by unitarity, and this gives us the bound
\es{NfNcbound}{
\Delta_{V_{\pm}} = N_f (1 - \Delta) - N_c + 1 \geq {1 \over 2} \,.
}
In the ${\cal N} = 4$ supersymmetric theory, where $\Delta$ is fixed to ${1 \over 2}$ because the R-symmetry is non-abelian, this constraint reduces to the ``good, bad, and ugly" classification of~\cite{Gaiotto:2008ak}. In that case theories with $N_f \geq 2 N_c$ were referred to as ``good" theories, in the sense that they have standard IR critical points, and theories with $N_f < 2 N_c - 1$ were referred to as ``bad" theories which do not satisfy the constraint.
The ``ugly" theories, for which $N_f = 2 N_c - 1$, have a free twisted hypermultiplet, containing $V_+$ and $V_-$, at their IR fixed points.  We will not comment any further on the ${\cal N} = 4$ theory.

In the ${\cal N}=2$ supersymmetric $U(N_c)$ theory, we may combine the monopole untiarity bound \eqref{NfNcbound} with the
meson operator unitarity bound, $\Delta> 1/4$, to get the constraint
\es{analytbound}{
N_f > \frac 4 3 N_c - \frac 2 3 \,.
}
We will work mostly in the Veneziano limit, which is defined by taking large $N_c$ with the ratio $x = {N_f \over N_c}$ held fixed.  The constraint in~\eqref{NfNcbound} becomes
\es{DeltaConv}{
\Delta \leq 1 - {1 \over x} \,.
}
We call the value of $x$ for which this bound is saturated $x_c$. Using \eqref{analytbound} we find the constraint $x_c> 4/3$, which implies that the standard electric $U(N_c)$ theory cannot be used all the way down to $x=1$. This immediately shows that there must be a ``crack in the conformal window.'' To find $x_c$ we need to calculate $\Delta$ as a function of $x$ in the large $N_c$ limit.  We address this calculation in the following subsection, but for now we quote the result $x_c \approx 1.45$.
Intriguingly, this is close to the value $3/2$ corresponding to the lower edge of the Seiberg superconformal window in 4 dimensions.

When $x < x_c$ a new global symmetry must appear in the IR which allows us to independently set $\Delta_{V_\pm} = 1/2$.  In section~\ref{emergent} we argue how these new global symmetries appear in Aharony's magnetic dual description of the theory.

\subsubsection{$\Delta$ and $F$ in the Veneziano limit
}

When $x > x_c$ we may use the localization procedure~\cite{Kapustin:2009kz,Jafferis:2010un} to calculate the scaling dimension $\Delta$ of the flavor superfields at the IR fixed point.  The $S^3$ partition function of the theory as a function of the trial R-charge $\Delta$ is
 \es{ZS3}{
 Z = {1 \over N_c ! }\int \left( \prod_{i = 1}^{N_c} { d \lambda_i \over 2 \pi} \right) \left( \prod_{i < j}^{N_c} 4\, \sinh^2 \left[ {\lambda_i - \lambda_j  \over 2} \right] \right) \prod_{i = 1}^{N_c} e^{N_f [ \ell(1 - \Delta + i {\lambda_i \over 2 \pi} ) + \ell(1 - \Delta - i {\lambda_i \over 2 \pi} ) ] } \,,
 }
 where
  \es{lz}{
 \ell(z) = - z \log\left(1 - e^{2 \pi i z} \right) + { i \over 2} \left( \pi z^2 + {1 \over \pi} \text{Li}_2 ( e^{2 \pi i z} ) \right) - {i \pi \over 12} \,.
 }
   The correct R-symmetry locally maximizes~\cite{Jafferis:2010un,Jafferis:2011zi,Closset:2012vg} the free energy $F = - \log \abs{Z}$.

 We may analyze the convergence of~\eqref{ZS3} by observing the integrand as the $\lambda_i \to \infty$.  For definiteness consider taking one of the integration variables $\lambda$ in the integrand to positive infinity, with the other integration variables held fixed.  We then use the expansions
 \es{ellExpand}{
 \ell\left(1 - \Delta \mp i {\lambda \over 2 \pi} \right) = \pm \,i\, {\lambda^2 \over 8 \pi} - {1 - \Delta \over 2} \lambda + O(\lambda^0)
 }
 and
 \es{sinhExpand}{
 \prod_{i < j}^{N_c} \sinh^2 \left[ {\lambda_i - \lambda_j  \over 2} \right] = e^{ (N_c - 1) \lambda + O(\lambda^0) } \,.
 }
 As noted in \cite{Willett:2011gp}, the partition function in~\eqref{ZS3} converges absolutely when
 \es{convergence}{
\Delta_{V_\pm} = N_f (1 - \Delta) - N_c + 1 > 0 \,.
 }
 In the Veneziano limit, this is indistinguishable from the unitarity bound~\eqref{NfNcbound}.

\begin{figure}[htb]
  \leavevmode
\begin{center}$
\begin{array}{cc}
\scalebox{.87}{\includegraphics{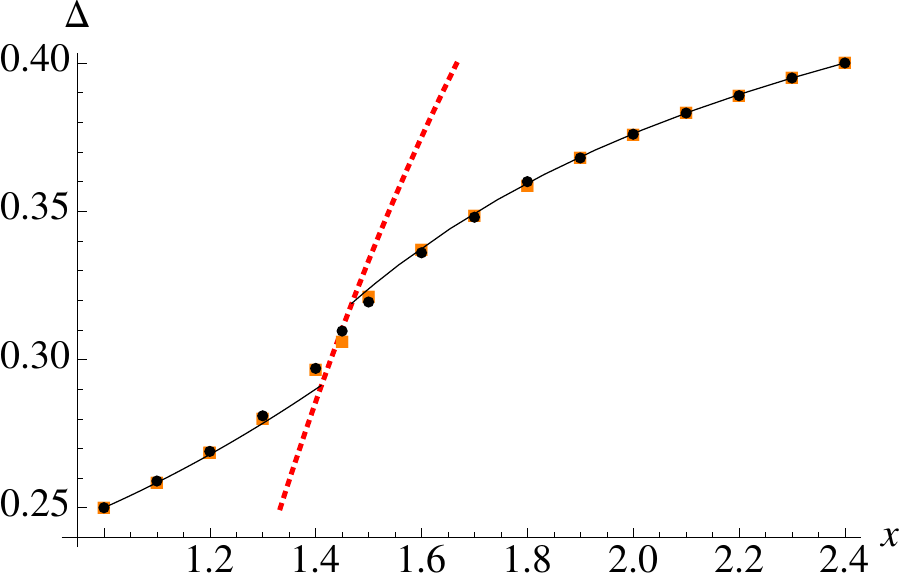}} & \scalebox{.87}{\includegraphics{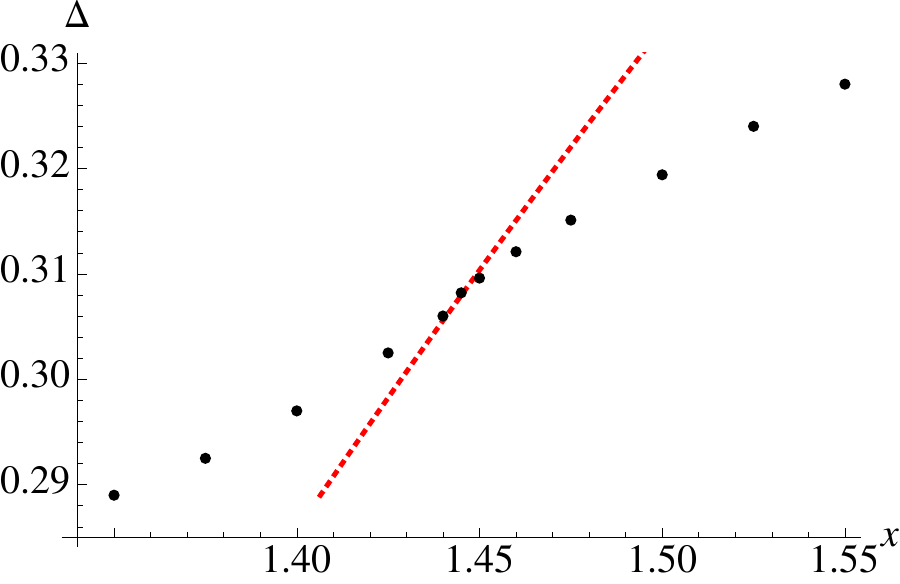}}
\end{array}$
\end{center}
\caption{ $\Delta$ as a function of $x = {N_f \over N_c}$ in the Veneziano limit.  The black points were computed using the saddle point method (method 1, section~\ref{Method1}) and the orange boxes were computed by extrapolating from the small $N_c$ numerical results (method 2, section~\ref{Method2}).  The dotted red curve is the convergence bound $1 - {1 \over x}$; we find that $\Delta$ meets the converge bound at the critical value $x_c \approx 1.45$.   The smooth black curves at large and small values of $x$ are the analytic approximations~\eqref{DeltaX} and~\eqref{DeltaXMag}, respectively. In the region right of the red curve we use the electric formulation of the theory, and in the region left of the red curve we use the magnetic formulation modified by decoupling the fields $V_\pm$.  The right plot is a zoomed in version of the left one and includes only the numerical results from method 1.  }
\label{DeltaXPlot}
\end{figure}

\begin{figure}[]
\begin{center}
\leavevmode
\scalebox{1.0}{\includegraphics{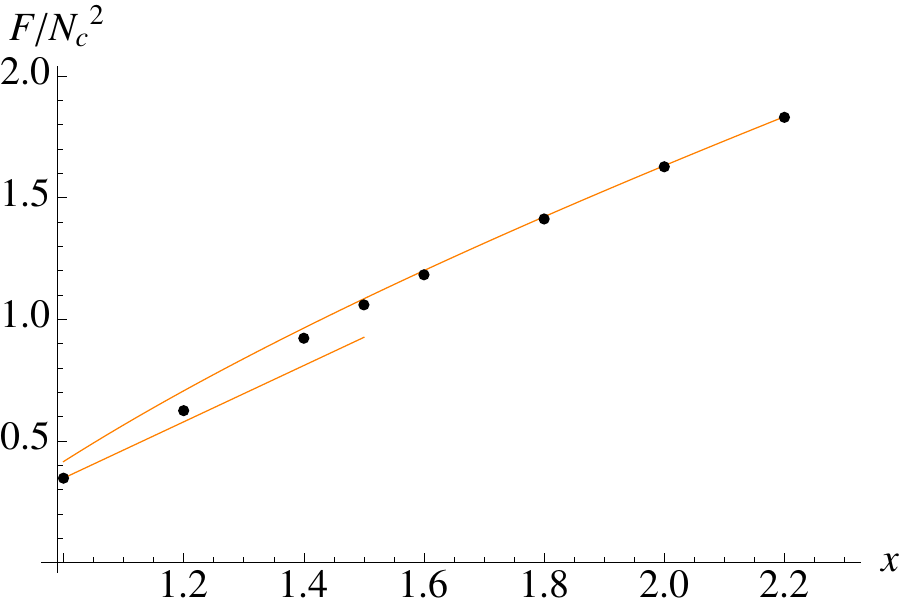}}
\end{center}
\caption{ $F/N_c^2$ in the Veneziano limit as a function of $x = {N_f \over N_c}$.  The free energy decreases monotonically as a function of $x$, consistent with the $F$-theorem.  The black points were computed numerically using the saddle point method (method 1, section~\ref{Method1}).  The upper orange curve is the analytic approximation~\eqref{FNcVeneziano} and the lower orange linear approximation at smaller $x$ is given in~\eqref{FNcVenezianoD}.      }
\label{FPlot}
\end{figure}

We compute $\Delta$ using three different methods, which are described more fully in section~\ref{Fmax}.  Method 1, described in section~\ref{Method1}, numerically computes the distribution of eigenvalues at the saddle point in the Veneziano limit using a procedure similar to that of \cite{Herzog:2010hf}.  Method 2 (section~\ref{Method2}) extrapolates small $N_c$ results, which can be computed by numerically integrating \eqref{ZS3}, to the Veneziano limit.  Method 3 (section~\ref{Method3}) computes $\Delta$ analytically in an asymptotic expansion in $1/x$, with the result
\es{DeltaX}{
\Delta(x) &= {1\over 2} - {2 \over \pi^2 x} + {2 ( 36 - 5 \, \pi^2 ) \over 3 \,\pi^4 x^2 } -  { 2 (\pi^2 - 12) (7 \, \pi^2 - 64) \over 3 \pi^6 x^3} \\
&-  {2  (47 \, \pi^6 - 1960 \, \pi^4 + 24960 \, \pi^2 - 100800 ) \over 15 \, \pi^8 x^4 } \\
& -  {2( 189 \, \pi^8 - 12832 \, \pi^6 + 289424 \, \pi^4 - 2679360 \, \pi^2 + 8847360) \over 45 \, \pi^{10} x^5} + O(1/x^6) \,.
}
In figure~\ref{DeltaXPlot} we plot the results for $\Delta$ as a function of $x$, and we determine numerically that $x_c \approx 1.45$.
In figure~\ref{FPlot}
we plot $F / N_c^2$ as a function of $x$, and we compare it to the asymptotic expansion
\es{FNcVeneziano}{
{F \over N_c^2} &= x \log 2 + {1 \over 2} \log x + \left( {3 \over 4} - \log 2 \right)  - {1 \over x} \left( {\pi^2-4 \over 2 \pi^2 } \right) \\
&- {1 \over x^2} \left( {512 - 112\, \pi^2 + 7 \, \pi^4 \over 24 \pi^4} \right) + O(1 / x^3 )  \,,
}
which we compute in section~\ref{Method3}.  The observation that $F / N_c^2$ is  a monotonically decreasing function of $x$ is consistent with the $F$-theorem~\cite{Jafferis:2011zi,Klebanov:2011gs,Myers:2010tj,Casini:2012ei}, since we may flow to theories with smaller $x$ by giving mass to some of the flavor multiplets.

\subsection{The Aharony duality} \label{AS}

We have found that the standard ``electric'' formulation of the $U(N_c)$ gauge theory does not work for
$x<x_c$. In particular, the $F$-maximization approach fails in this range because $F$ has no local maximum as a function of
the trial R-charge in the physically sensible range of $\Delta$.
The key to describing the IR fixed point of the theory with $x < x_c$ is to use the Aharony duality~\cite{Aharony:1997gp}.
The statement of the duality is that the IR fixed point of the ${\cal N} = 2$ theory with $U(N_c)$ gauge group and $N_f$ non-chiral flavors is dual to the IR fixed point of the ${\cal N} = 2$ theory with gauge group $U(N_f - N_c)$, $N_f$ non-chiral flavors $(q^a, \tilde q_a)$,
$N_f^2$ uncharged chiral multiplets $M^a_b$, and two uncharged chiral multiplets $V_+$ and $V_-$.  Additionally, the dual magnetic theory has the superpotential
\es{dualSuperPot}{
W = \tilde q_a M^a_b q^b + V_+ \tilde V_- + V_- \tilde V_+ \,,
}
where $ \tilde V_{\pm}$ are the monopole operators in the dual theory which create $\mp 1$ unit of flux through the $U(1)$ generated by the Cartan element $M = \text{ diag} (1, 0, \dots, 0)$.  Under the duality the monopole operators $V_{\pm}$ of the original theory are mapped to the gauge singlet chiral superfields of the dual theory.
 Similarly, the composite operators $\tilde Q^a Q_b$ of the original theory are mapped to the gauge singlet chiral superfields $M^a_b$ of the dual theory.  The field content of the dual theory along with the gauge and global symmetries are described in table~\ref{EMTableD}.
\begin{table}
\begin{center}
\begin{tabular}{| c | ccccc |}
 \hline
  Chiral Field  & $U(N_f - N_c)$ & $SU(N_f) \times SU(N_f)$ & $U(1)_A$ & $U(1)_J$ &  $U(1)_{R-\text{UV}}$ \\
   \cline{1-6} $\tilde q_a$ &  $N_f - N_c$ &  $(1, N_f)$  & $-1$ & $0$ & ${1 \over 2}$ \\
  $  q^a $ &  $\overline{N_f - N_c}$ &  $({\overline N_f} , 1)$  & $-1$ & $0$ & ${1 \over 2}$ \\
  $M^a_b$ & $1$ & $(N_f, \overline{N_f} )$ & 2 & 0 & 1 \\
 $ V_{\pm}$ & $1$ & $(1,1)$ & $-N_f$ & $\pm 1$ & $ {N_f \over 2} - N_c + 1$ \\
  $ \tilde V_{\pm}$ & $1$ & $(1,1)$ & $N_f$ & $\pm 1$ & $ -{N_f \over 2} + N_c + 1$ \\
  \hline
\end{tabular}
\end{center}
\caption{ The chiral superfields of the Aharony dual $U(N_f - N_c)$ SYM theory along with their charges under the gauge and global symmetries.  This theory is the magnetic dual of the $U(N_c)$ SYM theory with $N_f$ non-chiral flavors, with gauge and global symmetry assignments given in table~\ref{EMTable}.  Note that the $U(1)_R$ symmetry given is that of the UV fixed point. }
\label{EMTableD}
\end{table}
The duality has undergone many nontrivial checks (see, for example,~\cite{Aharony:1997gp,Willett:2011gp,Bashkirov:2011vy}).

\begin{figure}[]
\begin{center}
\leavevmode
\scalebox{.85}{\includegraphics{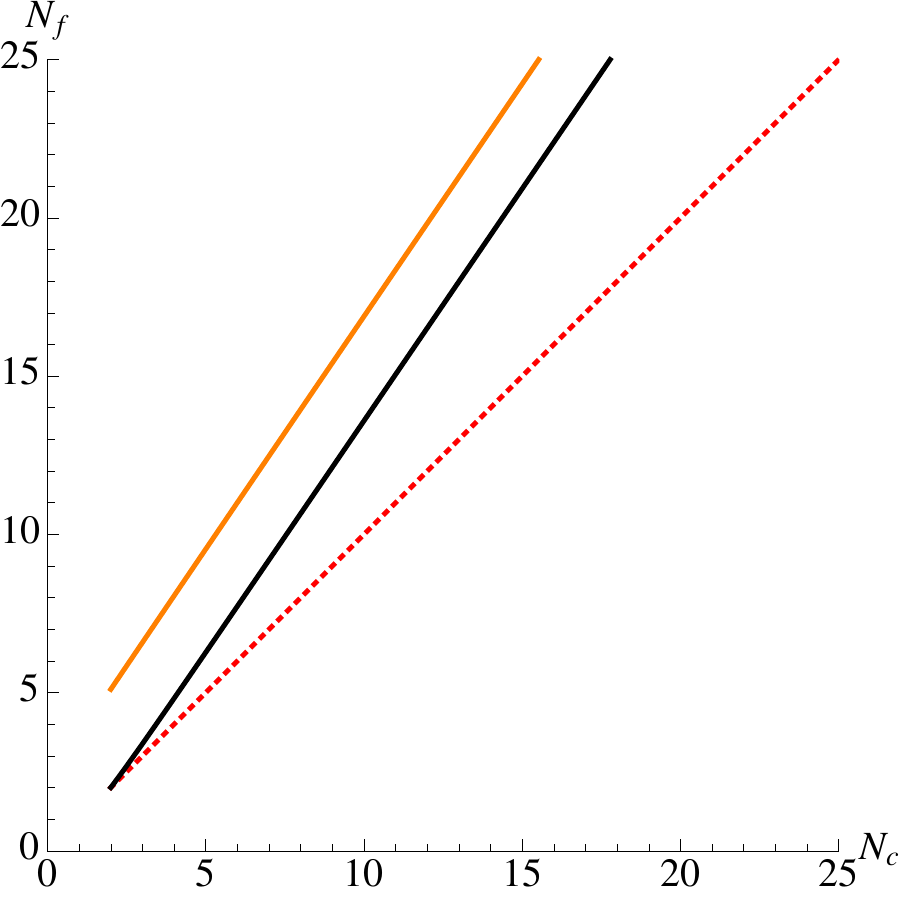}}
\end{center}
\caption{``The crack in the conformal window.'' The superconformal window in the $U(N_c)$ gauge theory with $N_f$ flavors is above the dotted red line $N_f = N_c$. The electric description has no emergent global symmetries above the black curve.
The ``standard'' magnetic localization prescription works between the black and orange curves.
These curves were calculated using the $1/N_f$ expansion result for $\Delta$ in~\eqref{DeltaFinal}.    }
\label{compare}
\end{figure}

The superpotential~\eqref{dualSuperPot} tells us that the dimension $\Delta_q$ of $q_a$ and $\tilde q^a$ is related to the dimension $\Delta$ of $Q^a$ and $\tilde Q_a$ in the original theory by $\Delta_q = 1 - \Delta$.  This allows us to write down the expression for
the R-charge of the monopole operator in the Aharony dual theory, analogous expression to~\eqref{NfNcbound}:
\es{NfNcboundDual}{
\Delta_{\tilde V_\pm}= N_c+1 - N_f (1-\Delta) \,.
}
Due to the superpotential \eqref{dualSuperPot}, the operators $\tilde V_\pm$ are not chiral primaries,
so the usual unitarity bound $\Delta_{\tilde V_\pm} \geq {1 \over 2}$ does not apply to them.
However, as we will see in the next section, the magnetic partition function on $S^3$ converges only when $\Delta_{\tilde V_\pm}>0$.
Thus, the ``standard'' localization prescription works in the magnetic theory in the narrow range
\es{narrowrange}{
-{1\over 2} \leq N_f (1-\Delta) - N_c < 1 \,
}
where $V_\pm$ are above the unitarity bound and the integral converges.
In the large $N_c$ limit with $x$ held fixed this range stays order $N_c^0$.  We may calculate this range in the $1/N_f$ expansion using the approximation for $\Delta$ in~\eqref{DeltaFinal} through order $1/N_f^5$.  A plot of the range (\ref{narrowrange})
in $(N_c, N_f)$ space is given in figure~\ref{compare}; the lower bound corresponds to the black curve and the upper bound to the orange curve.\footnote{Note that for each $N_f$ and $N_c$ there is unique $\Delta$ which locally maximizes $F$.  This implies, for example, that even though there is a range of $\Delta$ for which the magnetic theory partition function converges when $N_f$ and $N_c$ are taken above the orange curve in figure~\ref{compare}, we will not find a local maximum for $F$ in this range. }
For each $N_c>1$ there are a few values of $N_f$ that fall in the range (\ref{narrowrange}).

In the range (\ref{narrowrange}), i.e. between the black and orange curves in figure~\ref{compare}, the standard localization method works both in the electric and magnetic theories.
The computation in~\cite{Willett:2011gp} demonstrated the equality of the electric and magnetic partition functions in this range, providing evidence for the Aharony duality.  In the next section we will work outside of this range (below the black curve in~\ref{compare}), where the standard localization procedure does not apply in the electric theory.  We will assume the Aharony duality and extract information using the magnetic description
where we treat $V_\pm$ as decoupled free fields.

\subsection{An emergent global symmetry and $x < x_c$}  \label{emergent}

Going back to the Veneziano limit, the standard localization procedure fails in the electric theory when $x < x_c$.  The reason for this failure is the appearance of a new ``accidental'' global symmetry in the IR, which is related to the decoupling of the monopole fields $V_\pm$. This global symmetry is easily understood in the magnetic dual; it allows us to independently set the R-charges of $V_{\pm}$ to their free field value $1/2$, effectively decoupling these operators \cite{Agarwal:2012wd}. Said another way, the superpotential terms $W \supset V_+ \tilde V_- + V_- \tilde V_+ $ become irrelevant in the IR and can be ignored.
  Since the new global symmetry acts trivially on the dual gauge sector, we may use the localization procedure in the magnetic dual when $x < x_c$.

\subsubsection{$1 < x < x_c$} \label{1xlc}

When $1 < x < x_c$ we use the modified magnetic formulation of the theory, where the R-charges of $V_\pm$
are fixed to $1/2$.  The partition function of this theory as a function of the trial R-charge is
\es{Zdual}{
Z&= {e^{N_f^2 \ell(1- 2 \Delta) } \over 2 \, (N_f - N_c)! }  \int \left( \prod_{i = 1}^{N_f - N_c} { d \lambda_i \over 2 \pi} \right) \left( \prod_{i < j}^{N_f - N_c} 4\, \sinh^2 \left[ {\lambda_i - \lambda_j  \over 2} \right] \right) \\
&\prod_{i = 1}^{N_f - N_c} e^{N_f [ \ell(\Delta + i {\lambda_i \over 2 \pi} ) + \ell( \Delta - i {\lambda_i \over 2 \pi} ) ] } \,.
 }
 Note that even though the natural quantity in this theory is really the R-charge $\Delta_q$ of the dual fundamental fields, we perform all of our calculations in terms of $\Delta = 1 - \Delta_q$ to facilitate a comparison with the $x > x_c$ case.  The dual partition function converges absolutely when
$\Delta_{\tilde V_\pm}>0$.
  We compute $\Delta$ as a function of $x$ in the dual theory using the methods given in section~\ref{Fmax}, and the results are presented in figure~\ref{DeltaXPlot}.
   In the magnetic theory we may calculate analytic approximations to $\Delta$ and $F / N_c^2$ as asymptotic expansions in powers of $(x-1)$, with the results (see section~\ref{Method3})
  \es{DeltaXMag}{
  \Delta(x) = {1 \over 4} + {1 \over 4 \pi} (x - 1) + { (26 - 7 \, \pi) \pi - 8  \over 8 (\pi - 2) \pi^2 } (x-1)^2 + O\left( (x - 1)^3 \right)
  }
  and
  \es{FNcVenezianoD}{
  {F \over N_c^2} = {\log 2 \over 2} + (x-1) \left( {5 \log 2 \over 4} + {G \over \pi} \right) + O(x-1)^2 \,,
  }
where $G \approx 0.916$ is Catalan's constant.

  Note that $\Delta$ smoothly approaches $1/4$ as $x \to 1$ and also continuously connects with the electric curve at $x = x_c$.
  It appears that $\Delta$ and $F$ might be non-analytic at $x = x_c$, though our limited numerical precision does not allow us to make a precise statement. A general reason to expect a non-analyticity is that the global symmetry of the theory changes at $x_c$.

\subsubsection{$N_f = N_c$}

An interesting special case occurs at the boundary of the supersymmetric window, when $N = N_f = N_c$.
As shown in~\cite{Aharony:1997bx}, in this case there is an alternative description of the theory with chiral fields $V_+$, $V_-$ and $M^a_b = \tilde Q^a Q_b$ and superpotential
\es{superPot}{
W \sim V_+ V_- \det M\,.
}
The superpotential is everywhere non-singular, which means we may use it to describe the IR fixed point.
 When $N = 1$, the superpotential in~\eqref{superPot} is relevant, and, as shown in~\cite{Jafferis:2010un}, the theory flows to the IR fixed point where $V_+$, $V_-$, and $M$ have scaling dimensions ${2 \over 3}$.  When $N = 2$, the superpotential is marginal, but it was shown in~\cite{Willett:2011gp} that in fact the theory flows to the free theory in the IR where the fields $V_+$, $V_-$, and $M^a_b$ have scaling dimensions $1/2$.  When $N > 2$ the superpotential is irrelevant in the UV, so the theory trivially flows to the free theory in the IR.  However, as noticed in~\cite{Willett:2011gp}, the naive application of the localization procedure to the dual theory fails to reproduce this result.  The explanation for this apparent failure is simply that in these theories, which lie on the dotted red line in figure~\ref{compare}, a new global abelian symmetry appears in the IR which allows us to independently set the R-charges of $V_\pm$ to $1/2$.

 It is interesting to see how the last statement above is implemented at the level of the $S^3$ partition function for theories with $N > 2$. The partition function of the electric theory converges only for $\Delta<1/N$; for $N>3$ this excludes the values of
 $\Delta$ where the theory is unitary. The partition function of the naive magnetic dual, which includes the $V_{\pm}$ fields, is
 \es{MagDualNaive}{
 Z = e^{N^2 \ell(1- 2 \Delta) + 2 \ell(N\Delta) } \,.
 }
 The function $\ell(z)$ is defined in~\eqref{lz}, and it obeys the relation $\partial_z \ell(z) = - \pi z \cot( \pi z)$.  It is thus straightforward to check that for general $N > 2$, $Z$ is not extremized  by $\Delta = 1/4$ (in fact, for $N>3$ the electric integral does not even converge for this value of $\Delta$). For $N>2$ we do not find any local maximum of $F=-N^2 \ell(1- 2 \Delta) - 2 \ell(N\Delta)$ in the physically sensible range of $\Delta$.
  On the other hand, if we modify the magnetic theory by decoupling the monopole operators, then we may write
  \es{MagDual}{
 Z_{\rm modified} = e^{N^2 \ell(1- 2 \Delta) + 2  \ell(1 - \Delta_{\pm}) } \,.
 }
 We now see that, for all $N_f=N_c=N>1$, the partition function is minimized by $\Delta = 1/4$ and $\Delta_{\pm} =1/2$, as expected,
 and we have
\es{specialF}{
 F_{\rm edge} = \left ( \frac {N^2} {2} + 1\right ) \log 2\,.
 }

\subsection{RG flows and the $F$-theorem}

Table~\ref{numTab}
\begin{table}
\begin{center}
\begin{tabular}{| c | ccccccc |}
 \hline
    & $N_f = 1$ & $N_f = 2$ & $N_f = 3$ & $N_f = 4$ & $N_f = 5$ & $N_f = 6$ & $N_f = 7$ \\
   \cline{1-8} $N_c = 1$  & \spc{$1/3$\\$(.8724)$ } & \spc{$.4085$ \\$(1.934)$ } &  \spc{$.4370$ \\$(2.838)$ }  & \spc{$.4519$\\ $(3.679)$} &  \spc{$.4611$\\$(4.486)$ } & \spc{$.4674$\\$(5.272)$ } & \spc{$.4719$\\$(6.044)$ } \\
  $N_c = 2$ &  - & \spc{$1/4$\\$(2.079)$ }  & \spc{$.3417$\\$(4.722)$ } & \spc{$.3852$\\$(6.875)$ } &  \spc{$.4101$\\$(8.817)$ }& \spc{$.4263$\\ $(10.64)$ } & \spc{$.4375$\\ $(12.38)$} \\
  $N_c = 3$ & - & - & \fcolorbox{red}{white}{\spc{$1/4$\\$(3.812)$}} & \spc{$.3058$\\$(8.188)$} & \spc{$.3517$\\$(11.81)$} &  \spc{$.380$\\$(15.0)$} & \spc{$.400$\\$(18.0)$} \\
 $N_c = 4$ & - & - & - & \fcolorbox{red}{white}{\spc{$1/4$\\$(6.238)$}} & \spc{$.2809$ \\$(12.19)$}& \spc{$.3276$ \\$(17.51)$} & \spc{$.357$ \\$(22.2)$} \\
  $N_c = 5$ & - & - & - & - & \fcolorbox{red}{white}{\spc{$1/4$\\$(9.357)$}} & \fcolorbox{red}{white}{\spc{$.2672$\\$(16.62)$}} & \spc{$.3086$\\$(23.85)$} \\
  $N_c = 6$ & - & - & - & - & - & \fcolorbox{red}{white}{\spc{$1/4$ \\ $(13.17)$}} & \fcolorbox{red}{white}{\spc{$.2643$ \\ $(21.67$)}} \\
  $N_c = 7$ & - & - & - & - & - & - & \fcolorbox{red}{white}{ \spc{ $1/4$ \\ $(17.68)$}} \\
  \hline
\end{tabular}
\end{center}
\caption{ The scaling dimension $\Delta$ of the flavor multiplets and the value of $F$ (in parenthesis) at the conformal fixed points for a few small values of $N_f$ and $N_c$ in the ${\cal N} = 2$ SYM theory at vanishing Chern-Simons level. For $N_c$ and $N_f$
where we have to use the modified magnetic formulation of the theory with $V_\pm$ treated as free fields, (\ref{Zdual}), the results are enclosed in boxes.
  }
\label{numTab}
\end{table}
 summarizes our results for some small values of $N_c$ and $N_f\geq N_c$.
 We note that moving to the left along each row, which corresponds to RG flows associated with making some flavors massive,
 decreases the value of $F$ in agreement with the $F$-theorem \cite{Jafferis:2011zi,Klebanov:2011gs,Myers:2010tj,Casini:2012ei}.  This may also be seen in the Veneziano limit in figure~\ref{FPlot}. On the Higgs branch the theory may flow from $U(N_c)$ with $N_f$
 massless flavors in the UV to $U(N_c-1)$ with $N_f-1$
 massless flavors in the IR. According to table~\ref{numTab}, this movement along the diagonals makes $F$ decrease in agreement with the $F$-theorem.

\begin{figure}[]
\begin{center}
\leavevmode
\scalebox{1.0}{\includegraphics{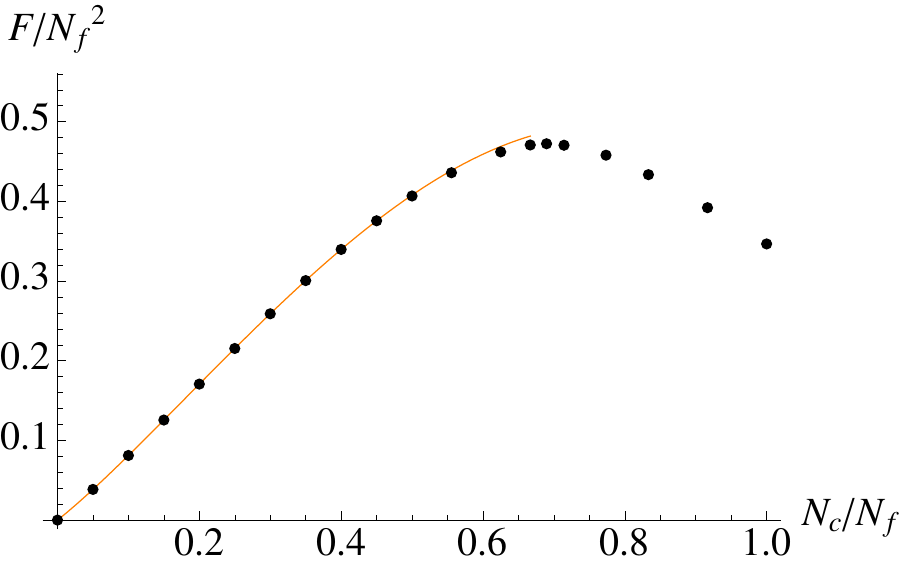}}
\end{center}
\caption{ $F/N_f^2$ in the Veneziano limit as a function of $ N_c / N_f$.  The quantity is peaked at the ``crack in the conformal window", $N_c / N_f \approx 1 / 1.45$.  The black points were computed numerically using the saddle point method (method 1, section~\ref{Method1}).  The left orange curve is calculated from the analytic approximation~\eqref{FNcVeneziano}, and the right orange curve at larger values of $N_c / N_f$ is calculated from~\eqref{FNcVenezianoD}.      }
\label{FPlot2}
\end{figure}

 We note, however, that $F$ is not monotonic along the columns. In moving down
 each column with $N_f>2$, $F$ first increases, peaks around the ``crack in the conformal window,'' and then decreases towards the edge
 of the window,
 $N_c=N_f$.
  This effect becomes more pronounced for large $N_f$.  In figure~\ref{FPlot2}
we plot $F/N_f^2$ in the Veneziano limit as a function of $N_c / N_f$, and we see that this quantity is peaked at the crack, $N_c / N_f  \approx 1/1.45$. Using figure~\ref{FPlot2}
 we estimate that near the crack the value of $F$ is $F_{\rm crack}\approx 0.47 N_f^2$.
 This is much bigger than the value at the edge $N_c=N_f$ from (\ref{specialF}), $F_{\rm edge}\approx 0.35 N_f^2$.

Let us propose a tentative interpretation of the non-monotonic behavior of $F$ at fixed $N_f$. In the standard electric $U(N_c)$
theory, on the Coulomb branch the gauge group may be broken to $U(N_c-1)\times U(1)$.
The $F$-theorem then tells us that $F_{U(N_c)} > F_{U(N_c - 1)}$ at fixed $N_f$. We observe this behavior for $N_c$ small enough that the monopole operators $V_\pm$, responsible for the Coulomb branch, are not decoupled; in the Veneziano limit this is the requirement $N_c \lesssim N_f / 1.45$.  For larger $N_c$ the monopole operators of the electric theory are decoupled, and
its Coulomb branch is no longer available. Instead, we can go on the Coulomb branch in the modified magnetic $U(N_f-N_c)$ theory,
and this {\it increases} $N_c$.
In this regime, which in the Veneziano limit corresponds to $N_c \gtrsim N_f / 1.45$, the $F$-theorem implies that $F$ is a decreasing function of $N_c$,
as may be observed in figure~\ref{FPlot2} and in table~\ref{numTab}.

Let us also note in passing that, in the 4-dimensional Seiberg conformal window, the Weyl anomaly coefficient
$a$ is not a monotonic function of $N_c$ at fixed $N_f$. The exact formula for $SU(N_c)$ gauge group with $N_f$ non-chiral flavors is \cite{Anselmi:1997am}
\es{exacta}{
 a = \frac {3} {16} \left ( 2 N_c^2- 1 - 3 \frac {N_c^4} {N_f^2} \right ) \ .
 }
In the Veneziano limit,
\es{exactaVen}{\frac {a} {N_f^2}= \frac {3} {16} \left ( 2 y - 3 y^2 \right )\ ,
}
where $y=(N_c/N_f)^2$. Clearly, $a /  N_f^2$ is maximized at $y=1/3$. This corresponds to $N_f=N_c \sqrt 3$,
which lies slightly above the strongly coupled edge of the conformal window, $N_f=3N_c/2$.

\section{Adding the Chern-Simons term}

It is instructive to understand what happens to the monopole operators when we turn on a Chern-Simons term at level $k$, which explicitly violates parity.
The Chern-Simons term gives a topological mass~\cite{Deser:1981wh} $m_T \sim g^2 k$ to the monopole operators, completely lifting the Coulomb branch of the theory.  Even though the monopole operators do not exist along the Coulomb branch, one may still worry that these operators exist at the origin of the moduli space.  This turns out not to be the case.  By looking at the Gauss law constraint, one sees that the monopole operators are not gauge invariant when $k \neq 0$.  To construct gauge invariant operators out of the monopole operators, one must act with flavor modes.  However, the resulting dressed monopole operators are no longer chiral~\cite{Bashkirov:2011vy}.\footnote{We thank Itamar Yaakov for helping to clarify this point.}  The matter content and global symmetries of the theory are the same as presented in table~\ref{EMTable}, except that the fields $V_{\pm}$ are no longer present.

When $\abs{k} > 0$ we may use the electric theory to describe the IR fixed point for the values of $N_f$ and $N_c$ which satisfy the condition
\es{SUSYCS}{
\abs{k} + N_f - N_c \geq 0
}
 for a supersymmetric vacuum. Now we do not have to worry about the dimensions of the monopole operators hitting the unitarity bound.
 The $S^3$ partition function nicely illustrates this observation.  The partition function of the $U(N_c)_k$ gauge theory at CS level $k$ with $N_f$ non-chiral flavors is given by
\es{ZS3CS}{
Z = {1 \over N_c!} \int \left( \prod_{i = 1}^{N_c} { d \lambda_i \over 2 \pi} e^{i k{ \lambda_i^2 \over 4 \pi}} \right) \left( \prod_{i < j}^{N_c} 4\, \sinh^2 \left[ {\lambda_i - \lambda_j  \over 2} \right] \right) \prod_{i = 1}^{N_c} e^{N_f [ \ell(1 - \Delta + i {\lambda_i \over 2 \pi} ) + \ell(1 - \Delta - i {\lambda_i \over 2 \pi} ) ] } \,.
}
When $k$ is non-zero we may always rotate the contour of integration so that~\eqref{ZS3CS} converges exponentially.

\subsection{The Giveon-Kutasov duality}

Magnetic dual descriptions of the IR fixed points are also available when we turn on Chern-Simons terms; they are described by the Giveon-Kutasov duality~\cite{Giveon:2008zn}.
The Giveon-Kutasov dual of the electric theory is similar to the Aharony dual of the $k = 0$ theory, whose matter content is given in table~\ref{EMTableD}, except that the fields $V_{\pm}$ and $\tilde V_{\pm}$ are no longer present.
The dual gauge group is $U(\abs{k} + N_f - N_c)_{-k}$, and the superpotential of the dual theory is simply $W = \tilde q_a M^a_b q^b$ ($ a,b = 1, \dots ,N_f$).  The $S^3$ partition function of the dual theory is
\es{ZdualCS}{
Z= {e^{N_f^2 \ell(1- 2 \Delta) } \over (\abs{k} + N_f - N_c )!}  \int& \left( \prod_{i = 1}^{\abs{k} + N_f - N_c} { d \lambda_i \over 2 \pi} e^{-i k{ \lambda_i^2 \over 4 \pi}}  \right) \left( \prod_{i < j}^{\abs{k}+N_f - N_c} 4\,\sinh^2 \left[ {\lambda_i - \lambda_j  \over 2} \right] \right) \\
&\prod_{i = 1}^{\abs{k} + N_f - N_c} e^{N_f [ \ell(\Delta + i {\lambda_i \over 2 \pi} ) + \ell( \Delta - i {\lambda_i \over 2 \pi} ) ] } \,.
 }
 We may use the magnetic formulation of the theory everywhere above the supersymmetry bound; on the three-sphere this gives
 results identical to the electric formulation \cite{Willett:2011gp}.
 The special case where $\abs{k} + N_f - N_c = 0$ is similar to the $k = 0$, $N_f = N_c$ case; the dual theory has no gauge group, and a simple calculation using~\eqref{ZdualCS} shows that $F$ is maximized at $\Delta = 1/4$.

 \subsection{Meson scaling dimensions}

It is instructive to keep track of the scaling dimension $2 \Delta$ of the gauge invariant mesons $M^a_b$ in the Veneziano limit.
 When we take the Veneziano limit in the theory with a CS level, we keep fixed both $\kappa = { \abs{k} \over N_c}$ and $ x ={N_f \over N_c}$.  We separately consider the regimes $0 < \kappa < 1$, $\kappa = 1$, and $\kappa > 1$.  In the first regime, when $0 < \kappa < 1$, the supersymmetry bound is given by $x = 1 - \kappa$, with $\Delta(x = 1 - \kappa) = {1 \over 4}$.  When $\kappa = 1$, the supersymmetry bound occurs when $x = 0$, and when $\kappa > 1$ we do not reach the supersymmetry bound for any value of $x$.
 The different regimes are considered separately below.

 \subsubsection{$\Delta$ when $0 < \kappa < 1$}

\begin{figure}[]
\begin{center}
\leavevmode
\scalebox{1.0}{\includegraphics{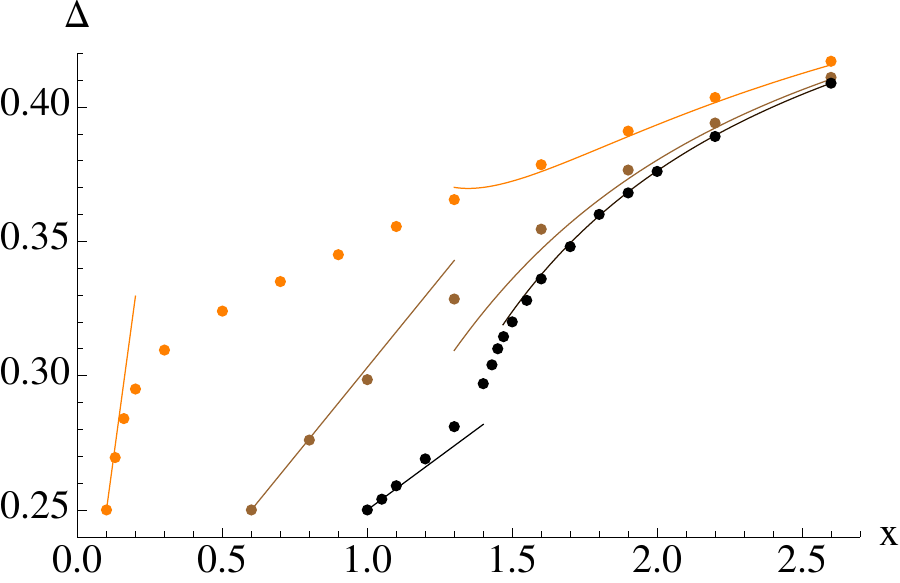}}
\end{center}
\caption{$\Delta$ as a function of $x = {N_f \over N_c}$ in the Veneziano limit at various value of $\kappa = \abs{k} / N_c$.  The black, brown, and orange points correspond to $\kappa = 0.01, \, 0.4, \, 0.9$, respectively.  The points were computed numerically using the saddle point method, described in section~\ref{Method1}.  The smooth curves at larger values of $x$ come from the analytic approximation to $\Delta$ in~\eqref{DeltaXCS}.  The linear approximations at small $x$ were plotted using the analytic approximation~\eqref{DeltaXCSMag}.    }
\label{DeltaXPlotCS}
\end{figure}

  We calculate $\Delta$ for various values of $0 < \kappa  < 1$ using the methods in section~\ref{Fmax}, with the results shown in figure~\ref{DeltaXPlotCS}.
We may use either the electric or magnetic formulations of the theory to calculate $\Delta$, and we verify numerically that they indeed give identical results.  As in the theory with $\kappa = 0$, we calculate analytic approximations to $\Delta$ about $x = \infty$ and $x = 0$ using the electric and magnetic descriptions of the theory, respectively.  The first few terms in the expansion about $x = \infty$ are
\es{DeltaXCS}{
\Delta(x,\kappa) = \Delta(x,\kappa = 0) + {8 \, \kappa^2 \over \pi^4 x^3} + { 16 \, (3 \pi^2 - 20) \kappa^2 \over \pi^6 x^4 } + O(1/x^5)\,,
 }
where $\Delta(x, \kappa = 0)$ is the result in~\eqref{DeltaX}, while the leading behavior in the expansion about $x = 0$ is given by
\es{DeltaXCSMag}{
\Delta(x,\kappa) =  {1 \over 4} + { x + \kappa - 1 \over 4 \pi ( 1- \kappa) } + O\left((x+\kappa - 1)^2 \right) \,.
}
An interesting observation, which may be seen in figure~\ref{DeltaXPlotCS}, is that $\partial_x \Delta$ diverges at small $x$ as $\kappa \to 1$ from below.  We will see in the following sections that $\Delta$ behaves qualitatively differently at small $x$ when $\kappa \geq 1$; in particular, when $\kappa \geq 1$ we find that $\Delta \geq {1 \over 3}$ for all $x$.

\subsubsection{The case $ \kappa = 1$}

\begin{figure}[]
\begin{center}
\leavevmode
\scalebox{1.0}{\includegraphics{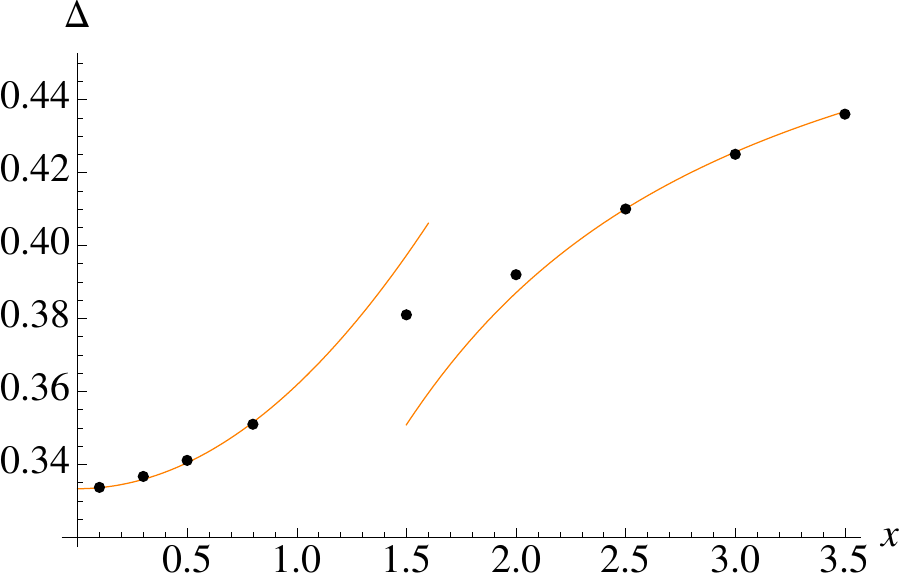}}
\end{center}
\caption{$\Delta$ as a function of $x = {N_f \over N_c}$ in the Veneziano limit with $\kappa = {\abs{k} \over N_c} = 1$.  The smooth orange curve at large $x$ was computed from the analytic approximation in~\eqref{DeltaXCS}, while the smooth orange curve at small $x$, which approaches $1/3$ at $x = 0$, comes from the analytic approximation in~\eqref{k1Dualpert}.  The black points were computed numerically using the saddle point method (method 1, section~\ref{Method1}).     }
\label{k1Figure}
\end{figure}

The theory with $\abs{k} = N_c$ is special since the magnetic dual is a $U(N_f)$ gauge theory at CS level $\mp N_c$, depending on whether $k = \pm N_c$, with $N_f$ non-chiral flavors $(q^a, \tilde q_a)$ and $N_f^2$ neutral mesons $M^a_b$; the rank of the dual gauge group does not grow with $N_c$.  Recall that this theory is also subject to the superpotential $W = \tilde q_a M^a_b q^b$.  It is interesting to analyze this theory in the limit $N_f \ll N_c$, since in this limit the large CS level makes the $U(N_f)$ gauge theory weakly coupled.
At the level of the partition function~\eqref{ZdualCS} this is seen by noting that in this limit the matrix integral localizes near the origin $\lambda_i = 0$, and the free energy reduces to
\es{partRed}{
F(\Delta)= N_f^2 \left( -\ell(1 - 2 \, \Delta) - 2 \,\ell(\Delta) + \frac 1 2 \log \frac {N_c}{N_f} + {\rm const} \right) + O(N_f^4/N_c^2) \,,
}
where
the logarithmic term comes from the
$U(N_f)_{N_c}$ supersymmetric Chern-Simons theory.
In the limit $N_f\ll N_c$ the free energy
is maximized by $\Delta = {1 \over 3}$.  As $x = {N_f \over N_c}$ is increased from zero to infinity, $\Delta$ is seen to increase monotonically from ${1 \over 3}$ to ${1 \over 2}$.  Using the methods of section~\ref{Method3}, we may solve perturbatively for $\Delta$ at small $x$ in the dual theory, with the result
\es{k1Dualpert}{
\Delta(x) = {1 \over 3} + {1 \over 324} \left( 99 - 20 \, \sqrt{3} \pi + { 81 ( 4 \, \sqrt{3} \, \pi - 9 ) \over 27 + 8 \, \pi (2 \, \pi - 3 \sqrt{3} ) } \right) x^2 + O(x^4) \,.
}
At large $x$ we may use the $1/x$ expansion from the electric theory in~\eqref{DeltaXCS}.  In figure~\ref{k1Figure} we plot these two analytic approximations along with the numerical results.

\subsubsection{Theories with $ \kappa > 1$}

We now discuss the theory with $\kappa > 1$ in the Veneziano limit.  The behavior of $\Delta$ as a function of $x = {N_f \over N_c}$ is qualitatively different at small $x$ from the behavior when $\kappa < 1$ and when $\kappa = 1$.  An important observation is that when $\kappa >1$ the scaling dimension $\Delta$ approaches ${1 \over 2}$ at small $x$, i.e. when $N_f$ is kept fixed while
$N_c$ and $k$ are sent to infinity.  This behavior is likely due to the higher spin symmetry in this limit~\cite{Aharony:2011jz,Giombi:2011kc,Chang:2012kt}.

\begin{figure}[]
\begin{center}
\leavevmode
\scalebox{1.0}{\includegraphics{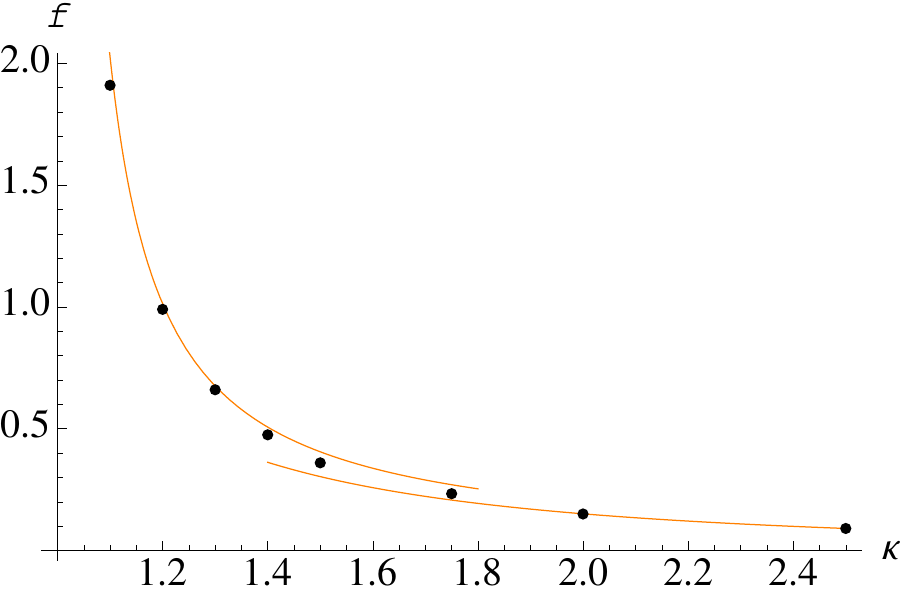}}
\end{center}
\caption{The function $f(\kappa)$, defined in~\eqref{DeltaXKappab1}, plotted over a range of $\kappa > 1$.  At values of $\kappa$ slightly greater than one, $f(\kappa)$ is well approximated by~\eqref{fkS}, which is the upper orange curve in the plot.  The lower orange curve is the approximation at large $\kappa$ given in~\eqref{fkL}.  The points were computed numerically using the saddle point method described in section~\ref{Method1}.    }
\label{fkPlot}
\end{figure}

The leading correction to $\Delta$ at small $x$ can be worked out in perturbation theory.  Writing
\es{DeltaXKappab1}{
\Delta(x,\kappa) = {1 \over 2} - x f(\kappa) + O(x^2) \,,
}
we can expand $f(\kappa)$ at large $\kappa$ by perturbation theory in the electric theory.  Using the methods in section~\ref{Method3} we find
\es{fkL}{
f(\kappa) = {1 \over 2 \, \kappa^2} + {\pi^2 \over 24 \, \kappa^4} + O(1/\kappa^6) \,.
}
At values of $\kappa$ near unity it is possible to approximate $f(\kappa)$ by perturbation theory in the magnetic theory, giving
\es{fkS}{
f(\kappa) = {2 \over \pi^2} {1 \over ( \kappa- 1)} + O \left( (\kappa - 1)^0 \right) \,,
}
which shows that $f(\kappa)$ diverges as $\kappa \to 1$ from above.  In figure~\ref{fkPlot}
we plot the analytic approximations to $f(\kappa)$ at small and large values of $\kappa$ along with numeric results.

Let us stress that our results for $\Delta$ in the large $N_c$ limit at fixed $N_f$ and $\lambda=N_c/k$ are {\it not} symmetric under $\lambda\rightarrow 1-\lambda$.
For small $\lambda$ we have \cite{Amariti:2011da}
\begin{equation}
\label{smalllambda}
\Delta_Q= {1\over 2}- {N_f N_c\over 2 k^2} + \ldots
\ ,
\end{equation}
while for $\lambda\rightarrow 1$
\begin{equation}
\label{smalllambda2}
\Delta_Q= {1\over 2}- {2 N_f \over \pi^2 (k- N_c)} + \ldots
\ .
\end{equation}
The lack of symmetry under $N_c\rightarrow k-N_c$ is due to the fact that the ${\cal N}=2$ Giveon-Kutasov duality does not relate
isomorphic theories; the electric theory has no superpotential, but the magnetic theory has a cubic superpotential.

\begin{figure}[]
\begin{center}
\leavevmode
\scalebox{1.0}{\includegraphics{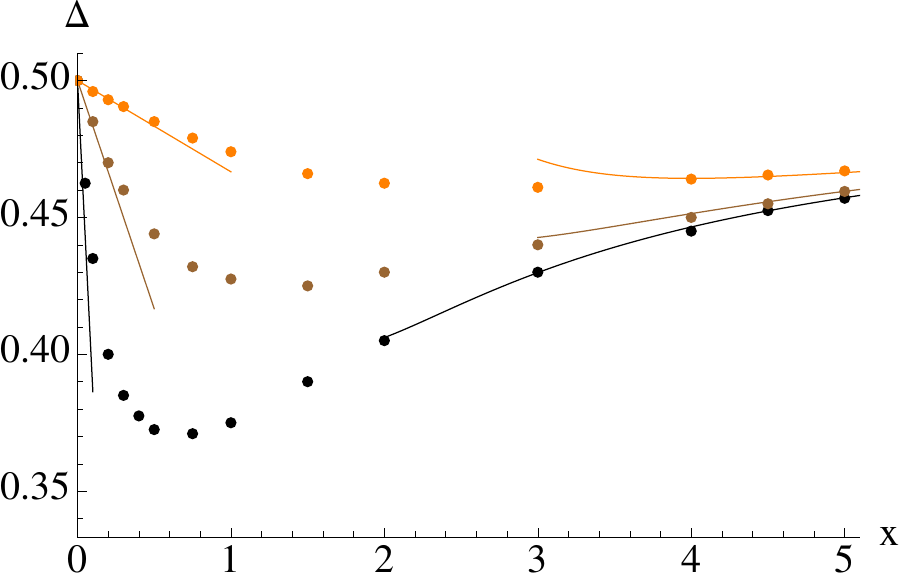}}
\end{center}
\caption{$\Delta$ as a function of $x = {N_f \over N_c}$ with $\kappa = {N_f \over N_c} = 1.2$ (black), $\kappa = 2$ (brown), and $\kappa = 4$ (orange).  The linear approximations at small $x$ were computed using~\eqref{DeltaXKappab1}, with $f(\kappa)$ plotted in figure~\ref{fkPlot}.  The analytic approximations at large $x$, which are shown as smooth curves, come from~\eqref{DeltaXCS}.  The points were computed numerically using the saddle point method described in section~\ref{Method1}.      }
\label{DeltaXPlotkb1}
\end{figure}

At large values of $x$ we may still use the analytic approximation to $\Delta$ in~\eqref{DeltaXCS}.
In figure~\ref{DeltaXPlotkb1}
 we plot $\Delta$ as a function of $x$ for a few different $\kappa > 1$ in order to illustrate the general behavior in this regime.  We also include the analytic approximations at small and large values of $x$ in their regimes of validity.

\section*{Acknowledgments}
We are very grateful to S. Pufu for collaboration during the early stages of this project. We thank S. Giombi, N. Seiberg, E. Witten and I. Yaakov for helpful discussions, and especially O. Aharony for his
important suggestions and comments on a draft of this paper. This research was supported in part by the US NSF under Grant No.~PHY-0756966.
IRK gratefully acknowledges support from the IBM Einstein Fellowship at the Institute for Advanced Study and from the John Simon Guggenheim Memorial Fellowship during his work on this paper. IRK is also grateful to the Aspen Center for Physics (NSF Grant No. 1066293) and
to the Simons Center for Geometry and Physics for hospitality.

\appendix

\section{$F$-maximization methods} \label{Fmax}

 The correct R-symmetry of the IR CFT locally maximizes the $S^3$ free energy \cite{Jafferis:2010un,Jafferis:2011zi,Closset:2012vg}.  The $S^3$ partition function may be calculated as a function of the trial R-charge using supersymmetric localization~\cite{Kapustin:2009kz,Jafferis:2010un}, which reduces the path integral to a finite number of ordinary integrals over the Cartan of the gauge group.
 In particular the integrals which we are interested in evaluating are given in~\eqref{ZS3}, \eqref{Zdual}, \eqref{ZS3CS}, and \eqref{ZdualCS}.
 Evaluating these integrals directly in the large $N_c$ limit is difficult when $N_f$ is of order $N_c$.  We take three approaches to approximating these integrals, and more importantly the critical R-charges, and we show that the different approaches give consistent results.  We will illustrate the methods explicitly for the theory without CS term since the generalizations to the cases with $k \neq 0$ are relatively straightforward.

 \subsection{Method 1:  The saddle point approximation} \label{Method1}

 In the large $N_c$ limit we may evaluate the integrals in the saddle point approximation; the integral localizes to the configuration of eigenvalues for which the integrand is an extremum.  Our first method numerically solves for this saddle point \cite{Herzog:2010hf}.  Since we only take into account the contribution from the saddle point, our approximation to the function $\Delta(x)$ is off by terms of order $1/N_c$.  We take into account finite $N_c$ corrections by repeating the calculation at increasing values of $N_c$ and then extrapolating to $N_c = \infty$.

 To begin it is instructive to rewrite the integral in~\eqref{ZS3} in the form
 \es{ZS32}{
 Z = {2^{ N_c (N_c - 1)} \over N_c!} \int \left( \prod_{i = 1}^{N_c} { d \lambda_i \over 2 \pi} \right) e^{-F[\lambda]} \,,
 }
 where
 \es{S1}{
 F[\lambda] &= - \sum_{i < j}^{N_c} \log \left( \sinh^2 \left[ { \lambda_i - \lambda_j \over 2 } \right] \right) \\
 &-  N_f  \sum_{i = 1}^{N_c} \left[ \ell\left(1 - \Delta + i {\lambda_i \over 2 \pi} \right) + \ell\left(1 - \Delta - i {\lambda_i \over 2 \pi} \right) \right] \,.
 }
 The saddle point configuration minimizes $F[\lambda]$,
 \es{actionMin}{
 { \partial F[\lambda] \over \partial {\lambda_i}} = 0 \,, \qquad i = 1, \dots, N_c \,,
 }
  and thus gives the dominant contribution to the partition function in the large $N_c$ limit.  The free energy $F = - \log \abs{Z}$ is then approximated by the real part of the functional $F[\lambda]$ evaluated on the saddle point.

\begin{figure}[]
\begin{center}
\leavevmode
\scalebox{1.0}{\includegraphics{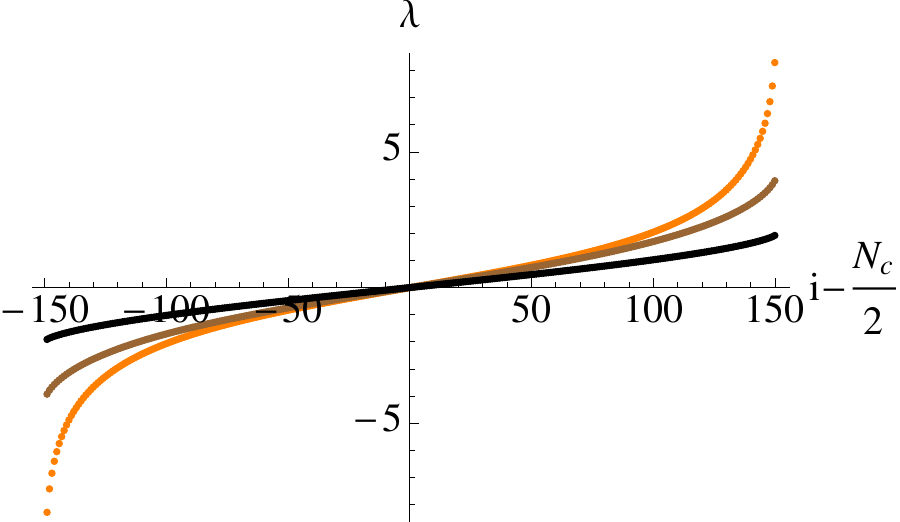}}
\end{center}
\caption{Eigenvalue distributions at the saddle point with $x = 1.5$ (orange), $2$ (brown), and $5$ (black), where $x = {N_f \over N_c}$.  We have taken $N_c = 300$ for this example.  The eigenvalues are manifestly antisymmetric about $ i = {N_c \over 2}$, where $i = 1, \dots , N_c$ labels the Cartan of the $U(N_c)$ gauge group.  As $x$ approaches the lower bound $x_c \approx 1.45$ the outer eigenvalues begin to repel each other.  }
\label{eigenDist}
\end{figure}

  We numerically solve the saddle point equations~\eqref{actionMin} following the prescription in~\cite{Herzog:2010hf}.  We solve the saddle point equations multiple times for each $x$, incrementing $\Delta$ each time, until we find the configuration which locally maximizes $F$.   Figure~\ref{eigenDist} shows a few eigenvalue distributions, computed at the critical $\Delta$, at increasing values of $x$ with $N_c = 300$.
As can be seen directly from~\eqref{S1}, the eigenvalues are antisymmetric about $i = {N_c \over 2}$.  The eigenvalues also remain order unity in the large $N_c$ limit, and as a result the free energy scales as $N_c^2$ at large $N_c$.  As $x$ approaches $x_c \approx 1.45$ the outer eigenvalues begin to repel each other.

\begin{figure}[]
\begin{center}
\leavevmode
\scalebox{1.0}{\includegraphics{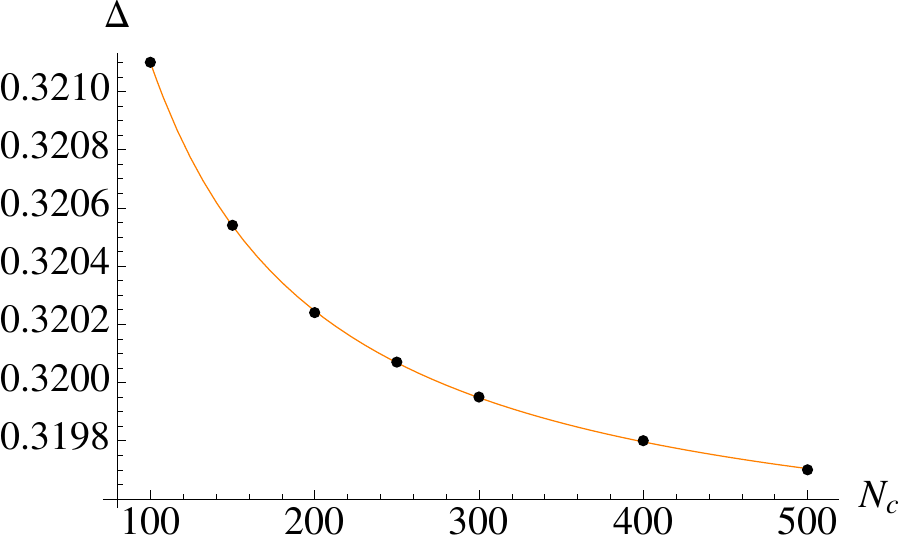}}
\end{center}
\caption{ A plot of $\Delta$ at $x = 1.5$ as a function of $N_c$ in the saddle point approximation, where $x = {N_f \over N_c}$.  The black points come from numerically solving the saddle point equations~\eqref{actionMin} and performing $F$-maximization.  The orange curve is a best fit to a function of the form~\eqref{DeltaNc}, with $\Delta_0 \approx .319$, $\Delta_1 \approx .189$, and $\Delta_2 \approx - 1.24$. }
\label{deltaNcPlot}
\end{figure}

For each $x$ we take into account finite $N_c$ corrections by computing $\Delta$ for a range of $N_c$ between $100$ and $500$ and fitting the results to a function of the form
\es{DeltaNc}{
 \Delta_0 + { \Delta_1 \over N_c} + {\Delta_2 \over N_c^2} + O(1/N_c^3) \,.
}
The quantity $\Delta_0$ is then a good approximation to the function $\Delta(x)$ at $N_c = \infty$.  We illustrate this procedure in figure~\ref{deltaNcPlot} for $x = 1.5$.
In that example we find $\Delta_0 \approx .319$.

 \subsection{Method 2: Extrapolating from small $N_c$} \label{Method2}

\begin{figure}[]
\begin{center}
\leavevmode
\scalebox{1.0}{\includegraphics{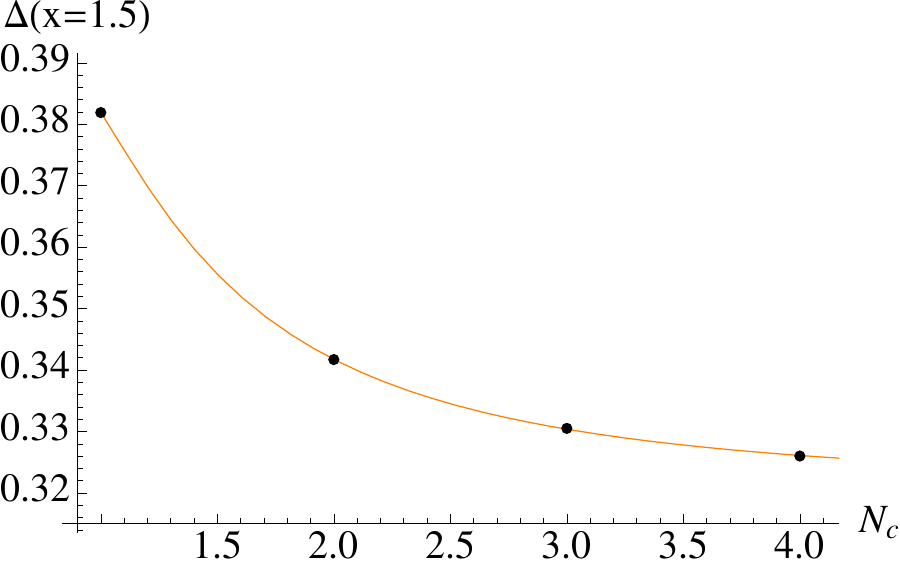}}
\end{center}
\caption{A plot of $\Delta$ at $x = 1.5$ as a function of $N_c$. The black points were computed by numerically integrating~\eqref{ZS3} for integer $1 \leq N_c \leq 4$.  The smooth orange curve fits these points to the expansion in~\eqref{DNcFcn}; in this example we find $\Delta_0 \approx .320$, $\Delta_1 \approx .0934 $, and $\Delta_2 \approx -.0319$.}
\label{DNcPlot}
\end{figure}

 Our second method numerically evaluates the partition function for $1 \leq N_c \leq 4$ over a range of $N_f$.  For each $N_f$ we find the $\Delta$ which locally maximizes the free energy.  We can then plot $\Delta$ at fixed $x$ as a function of $N_c$.  This is illustrated in figure~\ref{DNcPlot}, where we also include a fit to a function of the form
 \es{DNcFcn}{
 \Delta(x) = \Delta_0 (x) + {\Delta_1(x) \over N_c^2} + {\Delta_2(x) \over N_c^4} + O(1/N_c^6) \,.
 }
 The quantity $\Delta_0(x)$ should be the correct value for $\Delta(x)$ in the Veneziano limit.  The reason why the series in~\eqref{DNcFcn} only includes inverse powers of $N_c^2$ is explained in the following subsection.

 \subsection{Method 3:  The $1/x$ and $(x-1)$ expansions} \label{Method3}

 Our third method gives analytic approximations to $\Delta$ in the Veneziano limit at large $x$ and at $x$ slightly above unity.  These approximations are computed using the electric and magnetic formulations of the theory, respectively.

 \subsubsection{$1/x$ expansion in the electric theory}

    In the limit $N_f \gg N_c$, the effects of the gauge field are small and the flavors are almost free; $\Delta \approx {1 \over 2}$.  We perturbatively evaluate corrections to $\Delta = {1 \over 2}$ in a $1/x$ expansion through order $1/x^5$.  This expansion turns out to be a good approximation all the way down to $x \approx x_c$.

 We begin by reviewing the theory with $N_c = 1$ and $x = N_f$, since this example, while technically the simplest, still illustrates the main points of the $1/x$ procedure.  The $U(1)$ theory was discussed in some detail in~\cite{Klebanov:2011td}, so we will keep this discussion brief.  The partition function as a function of $\Delta$ is simply given by
  \es{Z1}{
 Z = \int_{-\infty}^{+\infty} { d \lambda \over 2 \pi} \,  e^{N_f [ \ell(1 - \Delta + i {\lambda \over 2 \pi} ) + \ell(1 - \Delta - i {\lambda \over 2 \pi} ) ] } \,.
 }
 In the limit of large $N_f$ we may evaluate~\eqref{Z1} in the saddle point approximation; the partition function localizes around $\lambda = 0$.  At $\lambda = 0$ the integrand in~\eqref{Z1} is minimized at $\Delta = {1 \over 2}$, which shows that $\Delta = {1 \over 2} + O(1/N_f)$.

 To calculate the higher order terms in the $1/N_f$ expansion we begin by writing
 \es{DeltaExpand}{
 \Delta = \Delta_0 + {\Delta_1 \over N_f} + {\Delta_2 \over N_f^2} + \cdots
 }
and rescaling the Cartan generator $\lambda$ to $\tilde \lambda= \lambda (2 \pi)^{-1} \sqrt{N_f}$.  Substituting~\eqref{DeltaExpand} into~\eqref{Z1}, we expand in powers of $1/N_f$ to obtain
\es{ZU1Exp}{
Z = {1 \over 2^{N_f} \sqrt{N_f} } \int_{-\infty}^{+\infty} d \tilde \lambda e^{- \pi^2 \tilde \lambda^2 / 2} \left[ 1 + {6 \pi^2 \Delta_1^2 + 24 \Delta_1 \tilde \lambda^2 + \tilde \lambda^4 \over 12 N_f } + \dots \right] \,.
}
We may perform the integrals in~\eqref{ZU1Exp} term by term in the $1/N_f$ expansion.  Doing so leads to the free energy
\es{FreeU1}{
F_{U(1)}(\Delta) = N_f \log 2 + {1 \over 2} \log{ {\pi N_f  \over 2} } - \left( {\pi^2 \Delta_1^2 \over 2} + 2 \Delta_1 + {1 \over 4} \right) {1 \over N_f} + \cdots \,.
}
Maximizing~\eqref{FreeU1} with respect to $\Delta_1$ leads to $\Delta_1 = - 2 / \pi^2$.
This result is reproduced in appendix~\ref{feyn} using Feynman diagram techniques.  The diagram techniques also show that this result has the simple generalization
\es{D1}{
\Delta_1 = - {2 \, N_c \over \pi^2 }
}
for arbitrary $N_c$.

It is useful to verify explicitly~\eqref{D1} for a few small $N_c$ using the $F$-maximization procedure.  Increasing the rank of the gauge group to $N_c =2$, the partition function is given by the double integral
  \es{Z2}{
 Z = 2 \int_{-\infty}^{+\infty} { d \lambda_1 d \lambda_2 \over (2 \pi)^2} \, \sinh^2\left( {\lambda_1 - \lambda_2 \over 2} \right)  \sum_{i = 1}^2 e^{N_f [ \ell(1 - \Delta + i {\lambda_i \over 2 \pi} ) + \ell(1 - \Delta - i {\lambda_i \over 2 \pi} ) ] } \,.
 }
As in the $U(1)$ case, we see here that in the large $N_f$ limit the free energy is locally maximized by $\Delta = 1/2 + O(1/N_f)$.  To calculate the subleading terms in $\Delta$, we rescale the $\lambda_i$ to $\tilde \lambda_i= \lambda_i (2 \pi)^{-1} \sqrt{N_f}$ and we calculate the partition function as a function of $\Delta$ in the $1/N_f$ expansion:
\es{ZU2Exp}{
Z &= {\pi^2 \over 2^{N_f-1} N_f} \int_{-\infty}^{+\infty} d \tilde \lambda_1 d \tilde \lambda_2 e^{- \pi^2  (\tilde \lambda_1^2 + \tilde \lambda_2^2) / 2} (\tilde \lambda_1 - \tilde \lambda_2)^2 \left[ 1 + \right. \\
&+ \left. {1 \over N_f} {\pi^2 \over 12} \left( 12 \Delta_1^2 + 4 (\tilde \lambda_1 - \lambda_2)^2 + 24 \Delta_1 (\tilde \lambda_1^2 + \tilde \lambda_2^2) + \pi^2 (\tilde \lambda_1^4 + \tilde \lambda_2^4) \right) + O(1/N_f^2) \right]   \,,
}
which leads to
\es{FU2Exp}{
F_{U(2)}(\Delta) &= (2 N_f) \log 2 + \left[ 2 \log \left( { N_f \pi \over 2} \right) - \log 2 \pi \right] \\
&- {1 \over N_f} \left( {7 \over 2} + \Delta_1 ( 8 + \pi^2 \Delta_1) \right) + O(1/N_f^2) \,.
}
Maximizing~\eqref{FU2Exp} with respect to $\Delta_1$ yields $\Delta_1 = - 4 / \pi^2$, consistent with~\eqref{D1}.
Carrying out this procedure for $N_c = 3$ gives
\es{NcE3}{
F_{U(3)}(\Delta) &= (3 N_f) \log 2 + \left[ {9 \over 2} \log\left( {N_f \pi\over 2 } \right) -  3 \log(2 \pi) - \log 2 \right] \\
 &- {1 \over N_f} \left( {51 \over 4} + \Delta_1 \big( 18 + {3 \over 2} \pi^2 \Delta_1\big) \right) + O(1/N_f^2) \,,
}
which is maximized by $\Delta_1 = - 6 / \pi^2$; again, this is consistent with~\eqref{D1}.

Iterating the procedure described above, we are able to solve for $\Delta$ as a function of $N_f$ and $N_c$ to arbitrary order in $1/N_f$.  Below we list the terms through order $1/N_f^4$,
\es{DeltaFinal}{
&\Delta(N_f,N_c) = {1 \over 2} - N_c  {2 \over \pi^2 N_f} + { -2 \pi^2 (5 N_c^2 - 2) + 72 \, N_c^2 \over 3 \pi^4 N_f^2} \\
&- N_c \left[ {14 \pi^4 (N_c^2 - 1) - 8 \pi^2 (37 N_c^2 - 18) + 1536 \, N_c^2 \over 3 \pi^6 N_f^3 } \right]  \\
&+{1 \over 15 \, \pi^8 N_f^4} \left[ -2 \pi^6 (47 N_c^4 - 65 N_c^2 + 18) + 80 \pi^4 (49 N_c^4 - 49 N_c^2 + 4) \right. \\
&\left. - 320 \pi^2 (156 N_c^4 - 83 N_c^2) + 201600 \, N_c^4 \right] + O(1/N_f^5) \,.
}
Notice that $\Delta_i$ is a polynomial in $N_c^2$ of order $i/2$ if $i$ is even and is a polynomial in $N_c^2$ of order $(i-1)/2$ times $N_c$ if $i$ is odd.  This observation may be verified diagrammatically.
The expression for $\Delta$ in~\eqref{DeltaFinal} simplifies in the Veneziano limit, with $x = N_f / N_c$ held fixed, and gives the result~\eqref{DeltaX}. It is also interesting to note that the corrections to $\Delta(x)$ proceed in powers of $1/N_c^2$.

We also find that the $N_c$ dependence of the free energy is given by
\es{FNcFinal}{
F &= N_c N_F \log 2 + \left[ {N_c^2 \over 2} \log(\pi N_f) - {1 \over 2} N_c (N_c - 1) \log (2 \pi) - \log(1! 2! \cdots (N_c - 1)!)  \right]  \\
&+ {N_c \over N_f} \left[ - \left({\pi^2-4 \over 2 \pi^2 } \right) N_c^2 + {1 \over 4} \right] \\
&+ {N_c^2 \over N_f^2}  \left[- \left( {512 - 112\, \pi^2 + 7 \, \pi^4 \over 24 \pi^4} \right) N_c^2 + \left( {7 \over 24} - {8 \over 3 \, \pi^2} \right) \right] + O(1/N_f^3) \,.
}
  In the Veneziano limit the expression for $F$ in~\eqref{FNcFinal} reduces to~\eqref{FNcVeneziano}. These $1/x$ expansions
  of $\Delta(x)$ and $F(x)$ are not convergent series, but rather asymptotic ones. Keeping the first few terms provides good numerical approximations at large $x$.

\subsubsection{$(x-1)$ expansion in the magnetic theory}

Let's now consider the limit where $x$ is slightly above $1$.  We know that $\Delta = {1 \over 4}$ at the SUSY bound ($x= 1$), so it is tempting to search for an asymptotic expansion of the form
\es{DeltaD}{
\Delta = {1 \over 4} + \tilde \Delta_1 (x - 1) + \tilde \Delta_2 (x - 1)^2 + \dots \,.
}
Indeed, we may find such an expansion by perturbation theory in the Aharony dual, which is described in section~\ref{AS}.

It is convenient to make the choice $N_f = N_c + 1$, so that the partition function of the dual theory is given by
\es{Zdual2}{
Z= {e^{( N_c+1)^2 \ell(1- 2 \Delta) } \over 2}  \int { d \lambda \over 2 \pi}   e^{( N_c+1) [ \ell(\Delta + i {\lambda \over 2 \pi} ) + \ell( \Delta - i {\lambda \over 2 \pi} ) ] }
 }
 and so that~\eqref{DeltaD} takes the form
 \es{DeltaD2}{
 \Delta = {1 \over 4} +{ \tilde \Delta_1 \over N_c}  + { \tilde \Delta_2 \over N_c^2} + \dots \,.
}
Substituting~\eqref{DeltaD2} into~\eqref{Zdual2} and rescaling $\lambda \to \lambda / \sqrt{N_c}$, we may expand about $N_c = \infty$ to find
\es{FexpMag}{
F &= - \log \abs{ Z} = N_c^2 \, { \log 2 \over 2} + N_c  \, \left( {5\,\log 2 \over 4} + {G \over \pi} \right) + {1 \over 2} \log N_c  \\
&+ \left( {\pi \, \tilde \Delta_1 \over 2} (1 - 2 \, \pi \, \tilde \Delta_1) + {3 \log 2 \over 4} + {G \over \pi}  +  \log ( \pi - 2) + {3 \log \pi \over 2}   \right) + O(1/N_c) \,.
}
 Performing $F$-maximization at this order then gives $\tilde \Delta_1 = 1 / (4 \pi)$.  Expanding $F$ to one more order in $1/N_c$ and performing $F$-maximization for $\tilde \Delta_2$ gives the result in~\eqref{DeltaXMag}.  A simple way to check this result is to repeat the analysis with $N_f = N_c + n$ for other small $n$.  In doing so we may also verify that $F$ is approximated by~\eqref{FNcVenezianoD} in the Veneziano limit for $x$ slightly greater than one.

\section{The anomalous dimension through Feynman diagrams} \label{feyn}

In this section we reproduce~\eqref{D1} through Feynman diagram techniques, following closely the methods given in~\cite{Klebanov:2011td,PhysRevB.77.155105}, in the abelian theory.

\subsection{The vector multiplet effective action}

In terms of component fields the Lagrangian of the abelian ${\cal N} = 2$ theory with $N_f$ non-chiral flavors is given by
\es{N=2Lagrangian}{
\CL &=  {1 \over 4g^2} F_{\mu\nu}F^{\mu\nu} + \bar{\lambda} \gamma^\mu \partial_\mu \lambda + \abs{\partial_\mu \sigma}^2 \\
&+ \sum_{a = 1}^{N_f} \left[ \abs{D_\mu q_a}^2 + \bar{\psi_a} \gamma^\mu D_\mu \psi_a + \sqrt{2}(q^\dagger_a \lambda \psi_a- \bar{\psi_a}\bar{\lambda}q_a) + \sigma \sigma q_a q^\dagger_a + \bar{\psi}_a \sigma \psi_a \right. \\
&\left.+  \abs{D_\mu \tilde{q}_a}^2 + \bar{\tilde{\psi}}_a \gamma^\mu D_\mu \tilde{\psi}_a + \sqrt{2}(\tilde{q}^\dagger_a \lambda \tilde{\psi}_a- \bar{\tilde{\psi}}_a \bar{\lambda}\tilde{q}_a)  +\sigma \sigma \tilde{q}_a \tilde{q}^\dagger_a + \bar{\tilde{\psi}}_a \sigma \tilde{\psi}_a \right] \,,
}
 with $(q_a, \psi_a)$ and $(\tilde q_a, \tilde \psi_a)$ the positively and negatively charged flavor multiplets, and $A_\mu$, $\lambda$, $\sigma$ the vector field, gaugino and scalar, respectively, in the vector multiplet.  The gauge covariant derivative $D_\mu$ acts on fields of charge $q$ by $D_\mu = \partial_\mu - i q A_\mu$.

The effective action for the vector multiplet can be constructed perturbatively in $1/N_f$ at large $N_f$ by integrating out the flavor multiplets. Following~\cite{Klebanov:2011td}, we write the Euclidean partition function as
\es{EffZ}{
Z \approx Z_0[q_a, \psi_a, \tilde{q}_a, \tilde{\psi}_a] \int DA_\mu D\lambda D\sigma e^{-S_{\textrm{eff}}[A_\mu] - S_{\textrm{eff}}[\lambda]  - S_{\textrm{eff}}[\sigma]}\,,
}
where to leading order in $N_f$
\es{effS}{
S_{\textrm{eff}}[A_\mu] &= -{1 \over 2}\int d^3 x \sqrt{g(x)} \int d^3 x' \sqrt{g(x')} A_\mu (x) A_\nu (x')\langle J^\mu(x)J^\nu(x')\rangle \,, \\
 S_{\textrm{eff}}[\lambda] &= -\int d^3 x \sqrt{g(x)} \int d^3 x' \sqrt{g(x')} \bar{\lambda}(x)\langle \bar{\eta}(x)\eta(x')\rangle \lambda(x) \,,\\
 S_{\textrm{eff}}[\sigma] &=  -{1 \over 2} \int d^3 x \sqrt{g(x)} \int d^3 x' \sqrt{g(x')} \sigma (x) \sigma (x')\langle Q(x)Q(x')\rangle \,,
}
and
\es{Currents}{
J^\mu(x) &= \bar{\psi}_a(x) \gamma^\mu \psi_a(x) + i\bar{q}_a(x)\partial^\mu q_a(x) - iq_a(x) \partial^\mu \bar{q}_a(x) +  \bar{\tilde{\psi}}_a(x) \gamma^\mu \tilde{\psi}_a(x)\\ &+  i\bar{\tilde{q}}_a(x)\partial^\mu \tilde{q}_a(x) - i\tilde{q}_a(x) \partial^\mu \bar{\tilde{q}}_a(x) \,,\\
\eta(x) &= i\sqrt{2} (q^\dagger_a(x) \psi_a(x) + \tilde{q}^\dagger_a(x) \tilde{\psi}_a)\,, \qquad \bar{\eta}(x)  = -i\sqrt{2} (\bar{\psi}_a(x) q_a(x) + \bar{\tilde{\psi}}_a(x) \tilde{q}_a(x)) \,,\\
Q(x) &= \bar{\psi}_a(x) \psi_a(x) + \bar{\tilde{\psi}}_a(x) \tilde{\psi}_a(x)\, .
}
To proceed we use the free field two-point functions
\es{2pt-ftn}{
\langle \bar{q}_a(x) q_b(0) \rangle &= {\delta_{a b} \over 4\pi \abs{x}} \,, \\
\langle \psi_a(x) \bar{\psi}_b(0) \rangle &= {i \over 4\pi} {\delta_{a b} \gamma^\mu x_\mu \over \abs{x}^3} \,,
}
with similar expressions holding for $\tilde{q}_a$ and $\tilde{\psi}_a$, to calculate the two-point functions of the currents.  Fourier transforming into momentum space we find
\es{Currents2}{
\langle J^\mu(k)J^\nu(-k) \rangle &= {N_f \over 4} \abs{k}\left(\delta_{\mu \nu} - {k_\mu k_\nu \over \abs{k}^2}\right) \,, \\
\langle \bar{\eta}(k)\eta(-k) \rangle &= {N_f  \over 4} {\gamma^\mu{k_\mu} \over \abs{k}} \,, \\
\langle Q(k)Q(-k) \rangle &=  {N_f  \over 8} \abs{k} \,.
}
In Lorentz gauge $\partial_\mu A^\mu  = 0$ this leads to the propagators
\es{Propagators1}{
\langle A_\mu (k) A_\nu (-k) \rangle &= {4\abs{k} \over N_f }  {1 \over k^2}\left(\delta_{\mu \nu} - {k_\mu k_\nu \over k^2}\right) \,, \\
\langle \bar{\lambda}(k)\lambda(-k) \rangle &= {4\abs{k} \over N_f}  {\gamma^\mu{k_\mu} \over k^2} \,, \\
\langle \sigma(k) \sigma(-k) \rangle &= {8\abs{k} \over N_f } {1 \over k^2} \,.
}

\subsection{One loop anomalous dimension}

We will calculate the anomalous dimension $\gamma=\Delta-1$ of the flavor multiplet fermions $\psi$ and $\tilde \psi$ perturbatively in $1/N_f$.  In the UV the free fermion propagator is simply $G_0(k) = \slashed{k} /k^2$, while at the IR interacting fixed point the propagator is given by $G(k) = \slashed{k} / k^{2 - 2\gamma}$, with $\gamma$ small.
At the IR fixed point $2 \gamma$ is equated with the coefficient of the $ \slashed{k} \textrm{log}(k)$ term in the 1PI correction to the propagator $\Pi(k)$;
\es{2Delta}{
2 \gamma = \left. \Pi(k) \right|_{ \slashed{k} \textrm{log}(k) } \,.
}

\begin{figure}[]
\begin{center}
\leavevmode
\scalebox{1.0}{\includegraphics{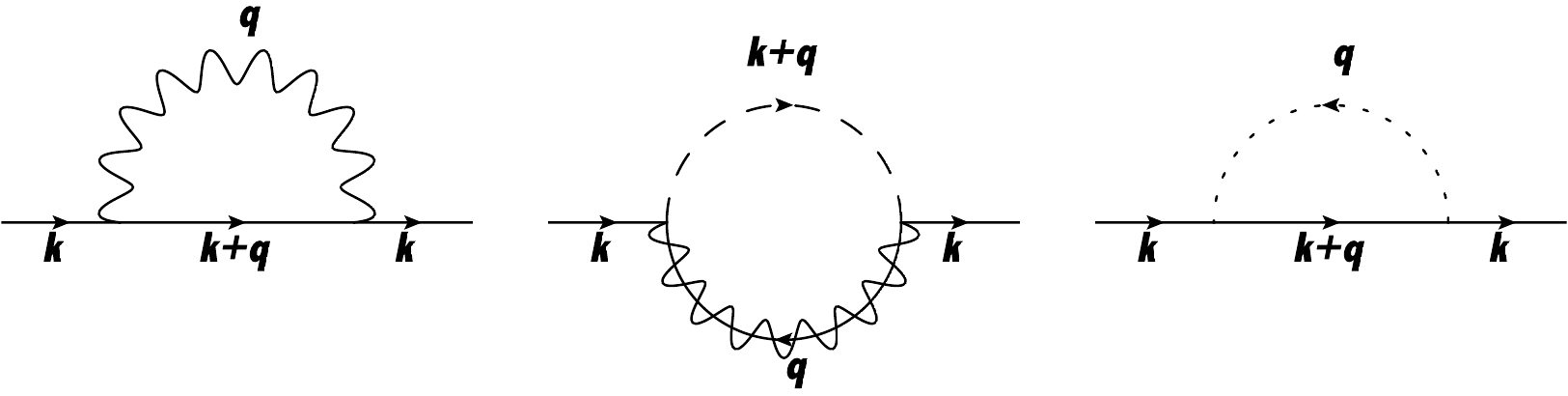}}
\end{center}
\caption{Self-energy diagrams for the fermion due to the gauge field (left, contributing $\Pi_A(k)$), gaugino (center, contributing $\Pi_\lambda(k)$) and neutral scalar (right, contributing $\Pi_\sigma(k)$). The wiggly line, solid line piercing wiggly line and coarse dotted line represent the gauge field, gaugino and neutral scalar, respectively.  The dotted line and solid line represent the flavor multiplet scalar and fermion.}
\label{selfenergy}
\end{figure}

As illustrated in figure~\ref{selfenergy},
at order $1/N_f$ there are three different contributions to $\Pi(k)$ coming from the gauge field, gaugino, and neutral scalar.
Denoting these corrections as $\Pi_A(k)$, $\Pi_\lambda(k)$, $\Pi_\sigma(k)$, respectively, we find
\es{fermion_anomalous_dim}{
\Pi_A(k) &=  \int {d^3 q \over (2\pi)^3} \gamma_\mu {\slashed{k}+\slashed{q} \over (k+q)^2} \gamma_\nu {4 \over N_f \abs{q}} \left(\delta_{\mu\nu} -{ q_\mu q_\nu \over q^2}\right) = - {4 \over 3 \, \pi^2 N_f} \slashed{k}\textrm{log}(k) + \cdots \,, \\
\Pi_\lambda(k) &=   \int {d^3 q \over (2\pi)^3} {1 \over q^2}  {\slashed{k}+\slashed{q} \over (k+q)^2} {4 \abs{k+q} \over N_f} = -{4 \over 3\pi^2 N_f }\slashed{k}\textrm{log}(k) + \cdots \,,\\
\Pi_\sigma(k) &=  \int {d^3 q \over (2\pi)^3} {1 \over q^2}  {\slashed{k}+\slashed{q} \over (k+q)^2} {8 q \over N_f} = -{4 \over 3\pi^2 N_f }\slashed{k}\textrm{log}(k)+\cdots \,,
}
where we have neglected terms whose $k$ dependence is not ${\slashed{k}} \log k$. Summing up all the corrections above we obtain the desired result~\eqref{D1}.

\bibliographystyle{ssg}
\bibliography{CGLP}

\end{document}